\documentclass[12pt]{article}
\usepackage{epsf}
\usepackage{latexsym,amssymb,euscript}
\usepackage[dvips]{graphicx}
\usepackage{amsfonts}
\usepackage{amsmath}
\begin{document}
\begin{center}

{\Large \bf Plaquette representation for $3D$ lattice 
gauge models: I. Formulation and perturbation theory}

\vspace*{0.6cm}
{\bf O.~Borisenko\footnote{email: oleg@bitp.kiev.ua}, \ 
S.~Voloshin\footnote{email: sun-burn@yandex.ru}} \\
\vspace*{0.3cm}
{\large \it
N.N.Bogolyubov Institute for Theoretical Physics, National Academy
of Sciences of Ukraine, 03143 Kiev, Ukraine}\\
\vspace*{0.5cm}
{\bf M.~Faber\footnote{email: faber@kph.tuwien.ac.at}} \\
\vspace*{0.3cm}
{\large \it Institut f\"ur Kernphysik,
Technische Universit\"at Wien}\\
\end{center}

\begin{abstract}
We develop an analytical approach for studying lattice gauge
theories within the plaquette representation where the plaquette
matrices play the role of the fundamental degrees of freedom.
We start from the original Batrouni formulation and show how it
can be modified in such a way that each non-abelian Bianchi
identity contains only two connectors instead of four. In addition, 
we include dynamical fermions in the plaquette formulation. 
Using this representation we construct the low-temperature perturbative 
expansion for $U(1)$ and $SU(N)$ models and discuss its uniformity 
in the volume. The final aim of this study is to give a mathematical 
background for working with non-abelian models in the plaquette formulation.
\end{abstract}

\section{Introduction}

\subsection{Motivation}

There exist several equivalent representations of lattice gauge
theories (LGT). Originally, LGT was formulated by K.~Wilson in terms
of group valued matrices on links as fundamental degrees of
freedom \cite{wilson}. The partition function can be written as 
\begin{equation}
Z = \int DU \ \exp \{ \beta S[U_n(x)] \} \ ,
\label{partfunc}
\end{equation}
\noindent
where $S[U_n(x)]$ is some gauge-invariant action whose naive 
continuum limit coincides with the Yang-Mills action. 
The integral in (\ref{partfunc}) is calculated over the Haar measure on 
the group at every link of the lattice. Very popular in the context of 
abelian LGT is the dual representation which was constructed 
in \cite{dualu1}-\cite{dualsun}. Extensions of dual formulations to 
non-abelian groups have been proposed only in the nineties in 
\cite{india}-\cite{conf4}. The resulting dual
representation appears to be a local theory of discrete variables which label
the irreducible representations of the underlying gauge group and can be written
solely in terms of group invariant objects like the $6j$-symbols, etc.
A closely related approach to the dual formulation is the so-called
plaquette representation invented originally in the continuum theory by
M.~Halpern \cite{halpern} and extended to lattice models in \cite{plrepr}.
In this representation the plaquette matrices play the role of the dynamical
degrees of freedom and satisfy certain constraints expressed through Bianchi
identities in every cube of the lattice.
Each representation has its own advantages and deficiencies. 
E.g., the Wilson formulation is well suited for Monte-Carlo
simulations, while dual and plaquette representations are usually used for 
an analytical study of the models. 
In particular, duals of abelian $U(1)$ LGT have been used to
prove the existence of the deconfinement phase transition at zero temperature
in four-dimensions ($4D$) \cite{u1dec,rigu1lgt} and to prove confinement at all 
couplings in $3D$ \cite{mack}. Also Monte-Carlo simulations proved to be very 
efficient in the dual of $4D$ $U(1)$ LGT \cite{zachdiss}.

So far, however both dual and plaquette formulations have not been so popular
in the case of non-abelian models, probably due to the complexity of these
representations. For instance, the plaquette representation can hardly be used
for Monte-Carlo computations due to a number of constraints on the plaquette
matrices. Let us remind the general form of the Batrouni construction . 
In \cite{plrepr} the plaquette representation was constructed in the maximal 
axial gauge. The partition function takes the following form if $S[U_n(x)]$ 
in (\ref{partfunc}) is the standard Wilson action 
\begin{equation}
\label{func}
Z = \int \prod_{p} dV_{p} \exp \left[ \beta \sum_{p}
{\rm Re \ Tr} V_{p} \right] \prod_{c} J(V_c) \ ,
\end{equation}
\noindent
where $V_{p}\in SU(N)$ are plaquette matrices, $DV_{p}$ is the invariant Haar
measure of $SU(N)$. The product over $c$ runs over all cubes of
the lattice. The Jacobian $J(V_c)$ is given by
\begin{equation}
\label{func2}
J(V_{c}) =  \sum_r d_r \chi_r \left ( V_c \right ) \ ,
\end{equation}
\noindent
where the sum over $r$ is a sum over all representations of $SU(N)$,
$d_r=\chi_r(I)$ is the dimension of the representation $r$. The last expression
is nothing but an $SU(N)$ delta-function which introduces certain constraint on
the plaquette matrices. This constraint is just the lattice form of the
Bianchi identity. The $SU(N)$ character $\chi_r$ depends on an ordered product
of $SU(N)$ matrices as dictated by the Bianchi identity. Its exact form will be
given in the next section. An important point is that in the non-abelian case the
resulting constraints on the plaquette matrices appear to be highly-nonlocal
and this fact makes an analytical study of the model rather difficult.
In particular, it has prevented so far the construction of any well controlled
and useful weak-coupling expansion.

A different plaquette formulation of $3D$ $SU(N)$ LGT has been proposed 
in \cite{rusakov}. It has  a local form and does not require gauge
fixing. Unfortunately, as we have found by explicit computations the model
proposed in \cite{rusakov} does not coincide with the Wilson LGT contrary to
the claim of \cite{rusakov}. Exact calculations on a $3^3$ lattice
show the equivalence between Wilson LGT and the model of \cite{rusakov} but
the equivalence is lost already for a $4\times 3^2$ lattice. The point is that
the constraints on the plaquette matrices in the model of \cite{rusakov} do not
match the non-abelian Bianchi identity as will be seen from our explicit 
calculations. Nevertheless, we have found that a certain decomposition of 
the lattice made in \cite{rusakov} can be useful in simplifying
Batrouni's original representation. In the next section we use
some ideas of \cite{rusakov} to reduce the number of connectors in constraints
on plaquette matrices from four to two per each cube of the lattice.
This simplifies the whole representation but still it is quite involved.

Let us also mention that there exists a plaquette
representation in terms of so-called gauge-invariant plaquettes \cite{ginvplaq}.
This representation does not require gauge fixing and can be formulated
both on finite and infinite lattices. It is quite possible to work also with this
formulation, all the methods developed in this paper can be straightforwadly 
extended to the model of gauge-invariant plaquettes. We have nevertheless found 
that the plaquette formulation obtained in the maximal axial gauge is simpler to 
handle, especially on finite lattices. Moreover, we shall explain how our 
formulation can be extended to periodic lattices. 
In addition to the previous works \cite{plrepr}, \cite{rusakov}, \cite{ginvplaq} 
we include also dynamical fermions in the plaquette formulation. 

In spite of the complexity of the plaquette representation, we think it has 
certain advantages compared to the standard Wilson representation. 
Some of them have been mentioned and elaborated in \cite{plrepr} and 
\cite{meanplaq}. Duality transformations, Coulomb-gas
representation, strong coupling expansion look more natural and simpler in
the plaquette formulation. It is also possible to develop a mean-field method
which is gauge-invariant by construction and is in better agreement with
Monte-Carlo data than any mean-field approach based on the mean-link method
\cite{meanplaq}.

Nevertheless we believe that the main advantage of this formulation, 
not mentioned in \cite{plrepr}, \cite{ginvplaq} 
and \cite{meanplaq} lies in its applications to
the low-temperature region. Let $V_p$ be a plaquette matrix in $SU(N)$ LGT.
The rigorous result of \cite{mackpetk} asserts that the
probability $p(\xi )$ that ${\mbox {Tr}}(I-V_p)\geq \xi$ is bounded by
\begin{equation}
p(\xi ) \leq {\cal O }(e^{-b\beta\xi}) \ , \ \beta\to\infty \ , \
b={\rm {const}}
\label{chestplaq}
\end{equation}
\noindent
uniformly in the volume. Thus, all configurations with
$\xi\geq O(\beta^{-1})$ are exponentially suppressed.
This is equivalent to the statement that the Gibbs
measure of $SU(N)$ LGT at large $\beta$ is strongly concentrated around
configurations on which $V_p\approx I$.
This property justifies expansion of
the plaquette matrices around unity when $\beta$ is sufficiently
large while there is no such justification for the expansion
of link matrices, especially in the large volume limit.
In particular, we think that replacing ${\rm Tr}V_p$ in the Gibbs
measure by a Gaussian distribution, in the region of sufficiently large
$\beta$ is a well justified approximation. In fact, all
the corrections to this approximation must be non-universal.

Actually, this is one of our motivations to construct a low-temperature,
i.e. large-$\beta$ expansion of gauge models using the plaquette representation.
The well-known problem of the standard perturbation theory (PT), 
i.e. whether PT is uniformly valid in the volume, 
can be shortly formulated in the following way. When
the volume is fixed, and in the maximal axial gauge the link matrices perform
small fluctuations around the unit matrix and the PT works very well producing 
the asymptotic expansion in inverse powers of $\beta$. However,
in the large volume limit the integrand becomes arbitrarily flat even in the 
maximal axial gauge. It means that in the thermodynamic limit (TL) the system
deviates arbitrarily far from the ordered perturbative state, so that no saddle
point exists anymore, i.e. configurations of link matrices are distributed
uniformly in the group space.
That there are problems with the conventional PT was shown in \cite{seiler},
where it was demonstrated that the PT results depend on the boundary 
conditions (BC) used to reach the TL.
Fortunately, even in the TL the plaquette matrices are close to unity,
the inequality (\ref{chestplaq}) holds and thus provides a basis for
the construction of the low-temperature expansion in a different and
mathematically reliable way. In this paper we develop such a weak-coupling
expansion both for abelian and non-abelian models.

This paper is organised as follows. In the next section we give our plaquette
formulation of $3D$ $SU(N)$ LGT. We work in maximal axial gauge and 
consider a model with arbitrary local pure gauge action. For fermions we choose 
either the Wilson or the Kogut-Susskind action. The plaquette representation will 
be formulated on a dual lattice for the partition function, 't Hooft and 
Wilson loops. In section 3 we construct the weak-coupling expansion for
the abelian model using the plaquette representation. We give a general expansion
for the partition function, calculate the zero-order generating functional 
and show how to compute corrections and expectation values of Wilson loops. 
Then, we extend the weak-coupling expansion to an arbitrary
$SU(N)$ gauge model. Here we give a general expansion of the Boltzmann factor,
explain how to treat the Bianchi constraints in the expansion, compute the
generating functional and establish some simple Feynmann rules. Finally, we
discuss some features of the large-$\beta$ expansion in non-abelian models.
Our conclusions are presented in section 4. Some computations are moved to 
the Appendices. In Appendix A we study the link Green functions which appear 
as the main building blocks of the expansion in the plaquette formulation.
In the Appendices B and C we give all technical details for the calculation 
of the free energy expansion for SU(N) models in the plaquette representation.

\subsection{Notations and conventions}

We work on a $3D$ cubic lattice $\Lambda\in Z^d$
with lattice spacing $a=1$, a linear extension $L$ and $\vec{x}=(x,y,z)$,
$x,y,z\in [0,L-1]$ denote the sites of the lattice. We impose either 
free or Dirichlet BC in the third direction and the periodic BC in other directions. 
Let $G=U(N), SU(N)$;
$U_l=U_n(x)\in G$, $V_p\in G$ and $DU_l$, $DV_p$ denote the Haar measure 
on $G$. $\chi_r(U)$ and $d(r)$ will denote the character and dimension 
of the irreducible representation $\{ r \}$ of $G$, correspondingly. We treat 
models with local interaction. Let $H[U]$ be a real invariant
function on $G$ such that
\begin{equation}
\mid H[U] \mid  \ \leq H(I)
\label{huleq}
\end{equation}
for all $U$ and coefficients of the character expansion of
$\exp (\beta H)$
\begin{equation}
C[r] \ = \ \int DU \ \exp \left ( \beta H[U] \right ) \chi_r(U)
\label{charexpdef}
\end{equation}
exist. Introduce the plaquette matrix as 
\begin{equation}
V_p\:=\:U_n (x)U_m (x+e_n)U_n^{\dagger}
(x+e_m)U_m^{\dagger}(x) \ ,
\label{plaqch}
\end{equation}
\noindent
where $e_n$ is a unit vector in the direction $n$. The action of the pure gauge 
theory is taken as 
\begin{equation}
S_g[U_n(x)] = \sum_{p\in\Lambda} H \left [ V_p \right ] \ .
\label{pgaction}
\end{equation} 
The action for fermions we write down in the form (colour and spinor indices 
are suppressed) 
\begin{equation}
S_q[\overline{\psi}_f(x),\psi_f(x),U_n(x)] = 
\frac{1}{2}\sum_{x,x^{\prime}\in \Lambda}\sum_{f=1}^{N_f} \ 
\overline{\psi}_f(x)A_f(x,x^{\prime};U_l)\psi_f(x^{\prime}) \ ,
\label{fermaction}
\end{equation}
\noindent
where 
\begin{equation}
A_f(x,x^{\prime};U_l) = M_f\delta_{x,x^{\prime}} +
\frac{1}{2} \sum_{n=1}^d  \left [ \delta_{x+e_n,x^{\prime}}\xi_n(x) U_n(x) + 
\delta_{x-e_n,x^{\prime}} \overline{\xi}_n(x^{\prime})
U_n^{\dagger}(x^{\prime}) \right ] \ .
\label{fermmatr}
\end{equation}
\noindent
We have introduced here the following notations 
\begin{equation}
M_f = m_f - rd \ , \ \xi_n(x) = r + \gamma_n \ , \ 
\overline{\xi}_n(x) = r - \gamma_n
\label{Wferm}
\end{equation}
\noindent
for Wilson fermions and 
\begin{equation}
M_f = m_f \ , \ \xi_n(x) = \overline{\xi}_n(x) =  
\eta_{n}(x) = (-1)^{x_{1}+x_{2}+...+x_{n -1}} 
\label{KSferm}
\end{equation}
\noindent
for Kogut-Susskind fermions. $m_f$ is mass of the fermion field, 
$N_f$ is the number of quark flavours
and $r$ is the Wilson parameter ($r=1$ is a conventional choice).

After integrating out fermion degrees of freedom the partition function 
of the gauge theory on $\Lambda$ with the symmetry group $G$ can be written as 
\begin{equation}
Z_{\Lambda}(\beta , m_f, N_f) = \int \prod_{x,n}dU_{n}(x) 
\times \exp \left \{ \beta \sum_{p \in \Lambda}H [V_p] + 
\sum_{f=1}^{N_f} {\rm  Tr} \ln A_f(x,x^{\prime};U_l)  \right \} \ ,
\label{partfuncorig}
\end{equation}
where the trace is taken over space, colour and spinor indices. 

\section{Plaquette formulation and expression for the Jacobian}

In this section we give our formulation of the plaquette representation for
$3D$ $SU(N)$ LGT. A short description of our procedure can be found in 
\cite{plaqform} for pure gauge theory and in \cite{plaqferm} for a theory with 
fermions. In subsection 2.1 we calculate the plaquette representation 
for the partition function on a dual lattice. In the second subsection we 
derive the plaquette representation for some observables.

\subsection{Expression for the Jacobian}

\begin{figure}[ht]
\centerline{\epsfxsize=6cm \epsfbox{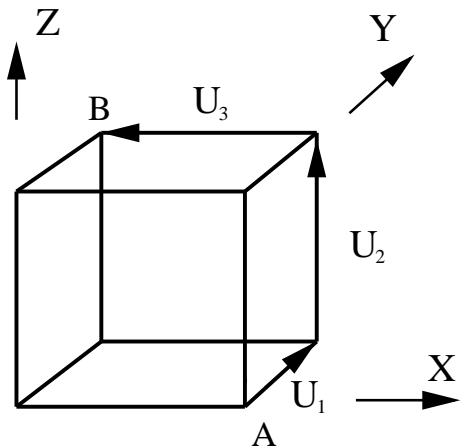}}
\caption{\label{cube} Vertices $A$, $B$ and the connector path on the cube.}
\end{figure}

\begin{figure}[ht]
\centerline{\epsfxsize=6cm \epsfbox{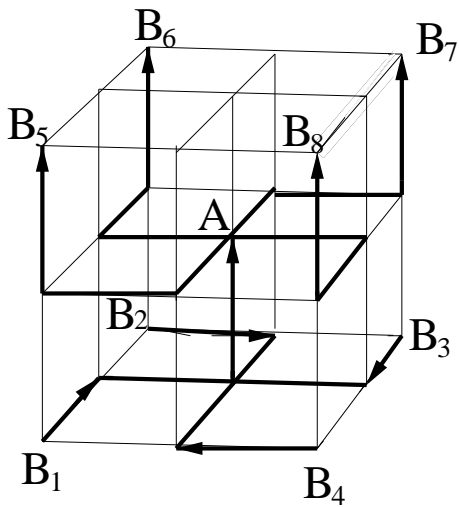}}
\caption{\label{connector} Structure of connectors on the lattice.}
\end{figure}

Let us first give a qualitative explanation of our transformations. 
Consider the partition function given in Eqs. (\ref{func}) and (\ref{func2}). 
For the formulation of the Bianchi identity 
related to a given $3d$-cube we define according to 
\cite{rusakov} a vertex $A$ in this cube and
separated by the diagonal a vertex $B$, see Fig.\ref{cube}. This assignment
we extend to all neighbouring cubes and finally to the whole lattice.
Thus, all $A$ vertices are separated by 2 lattice spacings,
the same is true for $B$ vertices. Next, we take a path connecting
the vertices $A$ and $B$ by three links $U_1$, $U_2$, $U_3$ as shown
in Fig.\ref{cube}. Then the matrix $V_c$ in (\ref{func}) and (\ref{func2}) 
entering the Bianchi identity for the cube $c$ can be presented 
in the following form
\begin{equation}
\label{ordprod}
V_{c} = \left ( \prod_{p \in A} V_p \right ) \ C
\left ( \prod_{p \in B} V_p \right ) \ C^{\dag} \ .
\end{equation}
$\prod_{p\in A}$ means an appropriately ordered product over three plaquettes
of the cube attached to the vertex $A$.
The matrix $C$ defines a parallel transport
of vertex $A$ into vertex $B$ and equals the product of three link matrices
connecting $A$ and $B$, see Fig.\ref{cube}
\begin{equation}
\label{func4}
C=U_{1} U_{2} U_{3}^{\dag} \ .
\end{equation}
\noindent
The connector $C$ plays a crucial role in the non-abelian Bianchi identity.
Its path has to fit to the choices 
of $\prod_{p\in A}$ and $\prod_{p\in B}$.
We choose the structure of the connectors as shown in Fig.\ref{connector}.
This $8$-cube fragment of the lattice is repeated through the whole lattice
by simple translation. As is seen from Fig.\ref{connector} there are
only four different types of connectors.
For example, the connector $B_1A$ is the same as $B_7A$. The collection of cubes
with the same type of connectors, e.g. $B_1A$ form on a dual lattice a  body
centred cubic (BCC) lattice with double lattice spacing.
There are four different sub-lattices of
this type corresponding to the four types of connectors in Fig.\ref{connector}:
$B_1A$, $B_2A$, $B_3A$ and $B_4A$, and therefore four different types
of Bianchi identities.

\begin{figure}[ht]
\centerline{\epsfxsize=10cm \epsfbox{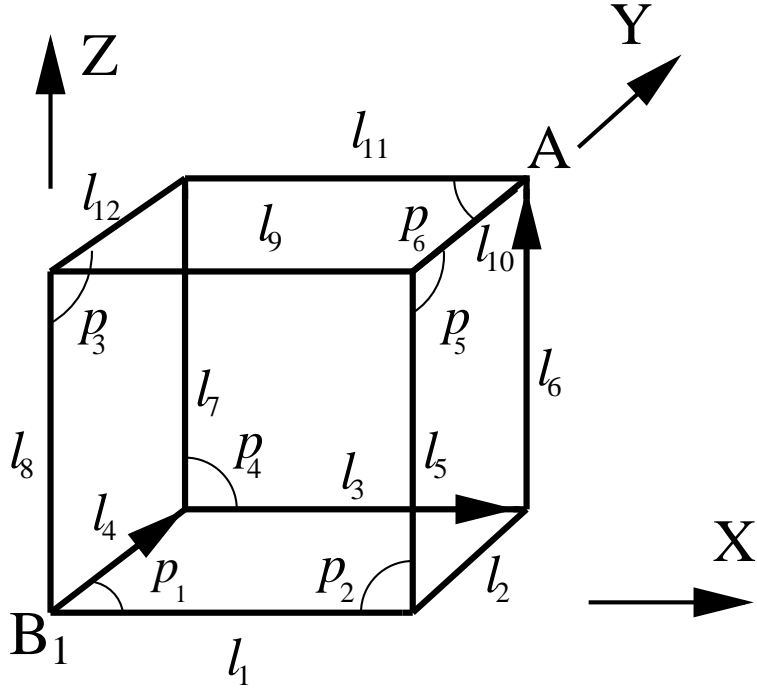}}
\caption{\label{origincube} Notations for links and plaquettes of
the original cube $B_1A$.}
\end{figure}

Consider now the partition function given by Eq. (\ref{partfuncorig}) on 
a $3D$ lattice with free or Dirichlet BC in the third direction 
and periodic BC in other directions for the link matrices.
To get the plaquette representation we make a change of variables (\ref{plaqch}) 
in the partition function (\ref{partfuncorig}).
Then the partition function gets the form 
\begin{equation}
Z_{\Lambda}(\beta , m_f, N_f) = \int \prod_pdV_p 
\exp \left \{ \beta \sum_{p \in \Lambda}H [V_p] \right \} 
\prod_p \ J(V_p) \ ,
\label{partfunctr}
\end{equation}
where the Jacobian of the transformation reads
\begin{equation}
J(V_p)\: =\: \int \prod_ldU_l \ \prod_p
\delta \left ( V_p^{\dagger} \prod_{l\in p}U_l - I \right ) 
\exp \left \{ \sum_{f=1}^{N_f} {\rm  Tr} 
\ln A_f(x,x^{\prime};U_l)  \right \} \ .
\label{jacobdef}
\end{equation}
\noindent
The last equation is rather formal, for we have to specify
the order of multiplications of non-abelian matrices.
An important point concerns the position of the plaquette matrix
within the product of link matrices. The plaquette matrices
$V_p$ we insert at the vertices $A$ or $B$ of each plaquette attached
to $A$ or $B$, correspondingly. To write down the plaquette delta-functions
in (\ref{jacobdef}) explicitly we use the following conventions:
1) a plaquette is always defined as
$U_n (x)U_m (x+e_n)U_n^{\dagger}(x+e_m)U_m^{\dagger}(x)$
with $n < m$, i.e.
it starts from the smallest possible coordinate; 2) plaquette matrices
are inserted as $V_p$, not $V_p^{\dag}$ in all vertices $A$ and $B$.
Then, for instance, the six delta-functions for the cube $B_1A$ in
Fig.\ref{connector} can be written with the notations of Fig.\ref{origincube} as
\begin{eqnarray}
\delta \left ( V_{p_1} U_{l_1}U_{l_2}U_{l_3}^{\dag}U_{l_4}^{\dag}
- I \right ) \ , \
\delta \left ( V_{p_2}U_{l_1}U_{l_5}U_{l_9}^{\dag}U_{l_8}^{\dag}
- I \right ) \ , \ \nonumber \\
\label{linkdelta}
\delta \left ( V_{p_3} U_{l_4}U_{l_7}U_{l_{12}}^{\dag}U_{l_8}^{\dag}
- I \right ) \ , \
\delta \left ( U_{l_3}U_{l_6}V_{p_4}U_{l_{11}}^{\dag}U_{l_7}^{\dag}
- I \right ) \ , \  \\
\delta \left ( U_{l_2}U_{l_6}V_{p_5}U_{l_{10}}^{\dag}U_{l_5}^{\dag}
- I \right ) \ , \
\delta \left ( U_{l_9}U_{l_{10}}V_{p_6}U_{l_{11}}^{\dag}U_{l_{12}}^{\dag}
- I \right ) \ . \nonumber
\end{eqnarray}
\noindent
The change of variables in (\ref{plaqch}) is uniquely determined from these
expressions. Similar expressions can be written for all types of connectors
using the same rules. Now we want to map one of the plaquette delta's
to a delta-function for the cube using all other delta's.
This procedure can be done for all cubes of the lattice in the following way.
We start from the plane $z=L-1$ and use space-like plaquettes
(i.e., lying in the $xy$-plane) to map them onto cube delta-function.
For the above cube $B_1A$ it means that we take the plaquette delta for $V_{p_6}$
and substitute link matrices by their expressions obtained from all other delta's.
Then we repeat these calculations for all cubes.
One should be careful about the order in which the cubes
are taken since to accomplish such calculations for all cubes, the order is important.
We can proceed strictly plane by plane and stop at the plane $z=1$.
There are still space-like plaquettes left in the plane $z=0$.
The connector for a given cube is taken in accordance with Fig.\ref{connector}.
Therefore, we mapped all space-like plaquette delta's lying on the top
of a given cube onto cube deltas with connectors lying on the bottom of this cube
(in the gauge $U_3=I$ defined below). For instance, if the original plaquette lies in
the plane $z=z_i$, then the connector lies in the plane $z_i-1$.
These cube delta-functions become samples for the Bianchi constraints.
In this way we arrive at the following expression for the Jacobian (\ref{jacobdef})
\begin{eqnarray}
J(V_p)\: =\: \int \prod_ldU_l \ \prod_{p_0} \delta \left ( V_{p_0}^{\dagger}
\prod_{l\in p_0}U_l - I \right ) \prod_{i=1}^4 \prod_{c(i)} J(V_c^{(i)})
\nonumber   \\
\times \exp \left \{ \sum_{f=1}^{N_f} {\rm  Tr} 
\ln A_f(x,x^{\prime};U_l)  \right \} \ ,
\label{jacob1}
\end{eqnarray}
\noindent
where $J(V_c)$ is given in (\ref{func2}). $\prod_{p_0}$ runs over all
time-like plaquettes and $\prod_{c(i)}$ means a product over all cubes with
the $i$-th type of connector. In the unique notations of Fig.\ref{origincube},
the matrices $V_c^{(i)}$ for different types of connectors $C_{B_iA}$
(see Fig.\ref{connector}) are given by
\begin{equation}
V_c^{1} = V_{p_3}^{\dag}V_{p_2}V_{p_1}^{\dag}C_{B_1A}V_{p_5}V_{p_6}
V_{p_4}^{\dag}C_{B_1A}^{\dag} \ , \ C_{B_1A} = U_{l_4}U_{l_3}U_{l_6} \ ,
\label{cdB1A}
\end{equation}
\begin{equation}
V_c^{2} = V_{p_4}^{\dag}V_{p_3}^{\dag}V_{p_1}^{\dag}C_{B_2A}V_{p_2}V_{p_6}
V_{p_5}C_{B_2A}^{\dag} \ , \ C_{B_2A} = U_{l_3}U_{l_2}^{\dag}U_{l_5} \ ,
\label{cdB2A}
\end{equation}
\begin{equation}
V_c^{3} = V_{p_5}V_{p_4}^{\dag}V_{p_1}^{\dag}C_{B_3A}V_{p_3}^{\dag}V_{p_6}
V_{p_2}C_{B_3A}^{\dag} \ , \ C_{B_3A} = U_{l_2}^{\dag}U_{l_1}^{\dag}U_{l_8} \ ,
\label{cdB3A}
\end{equation}
\begin{equation}
V_c^{4} = V_{p_2}V_{p_5}V_{p_1}^{\dag}C_{B_4A}V_{p_4}^{\dag}V_{p_6}
V_{p_3}^{\dag}C_{B_4A}^{\dag} \ , \ C_{B_4A} = U_{l_1}^{\dag}U_{l_4}U_{l_7} \ .
\label{cdB4A}
\end{equation}

\begin{figure}[ht]
\centerline{\epsfxsize=10cm \epsfbox{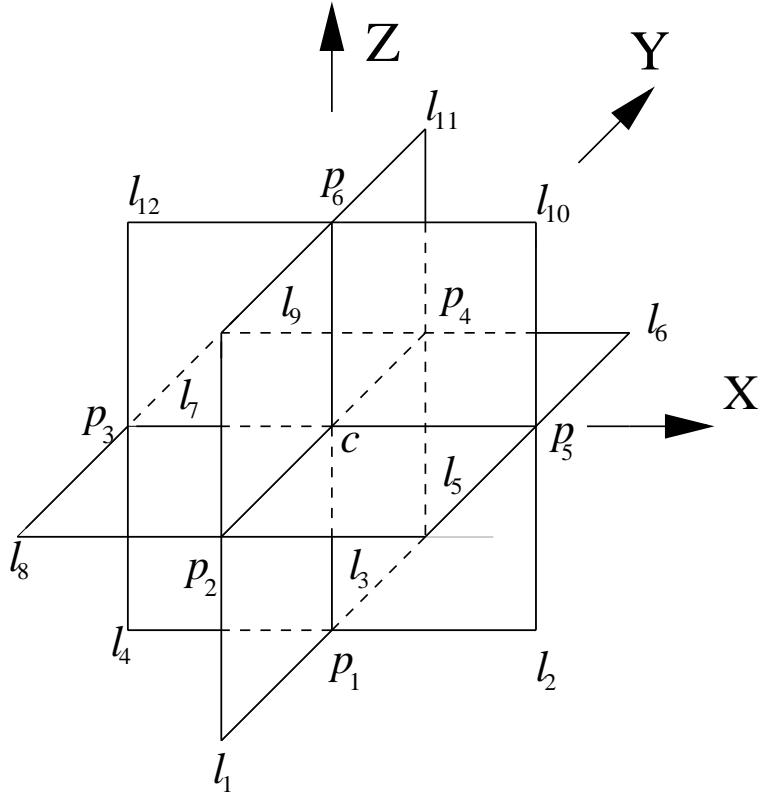}}
\caption{\label{dualcube} Original cube $B_1A$ on the dual lattice.}
\end{figure}

The rest of the derivation follows closely Ref.\cite{plrepr}.
Choose the maximal axial gauge in the form
\begin{equation}
\label{func5}
U_{3}(x,y,z)=U_{2}(x,y,0)=U_{1}(x,0,0)=I \ .
\end{equation}
\noindent
For the Dirichlet BC one has $U_{2}(x,y,0)=U_{1}(x,y,0)=I$. 
Note, that the gauge fixing procedure and the procedure described above are freely
interchangeable. After gauge fixing one can easily obtain expressions for link
variables in terms of plaquettes. First of all, space-like links from the
plane $z=L-1$ can be trivially integrated out removing all plaquette but
Bianchi delta's from the corresponding cubes near the top boundary of the lattice.
Further, we observe that expressions for link variables will depend
on the position of a link relatively to the vertex $A$ or $B$. As example,
consider a line containing vertices $B$ in $z$ direction, $x,y$ being fixed and
let one of the vertices $B$ have the coordinates $B=(x,y,z)$. Then one finds
for the link $l=(x,y,z;n_2)$
\begin{equation}
U_{n_2}(x,y,z)=\prod_{z^{\prime}=z-1}^0 V_{23} (x,y,z^{\prime}) \ ,
\label{linku1}
\end{equation}
\noindent
while for the link $l=(x,y-e_2,z;n_2)$
\begin{equation}
U_{n_2}(x,y-e_2,z)=\prod_{z^{\prime}=0}^{z-1}V_{23} (x,y-e_2,z^{\prime}) \ ,
\label{linku2}
\end{equation}
\noindent
where $V_{23}$ is the plaquette matrix from the $(zy)$-plane. 
Similar expressions one obtains for all links entering in or coming out
of the considered line. Obviously, the same rules apply for links in direction $n_1$
as well as for lines containing vertices $A$.
Using these expressions one can easily integrate out all link variables in
(\ref{jacob1}). Clearly, the presence of the fermion determinant cannot 
change the integration procedure. 
In what follows we pass to the dual lattice where cubes become
sites, links become plaquettes and sites become cubes. 
On the dual lattice our original notations
of Fig.\ref{origincube} for the cube $B_1A$ become as in  Fig.\ref{dualcube}.
This helps us to obtain explicit expressions for the Jacobian.

\begin{figure}[ht]
\centerline{\epsfxsize=12cm \epsfbox{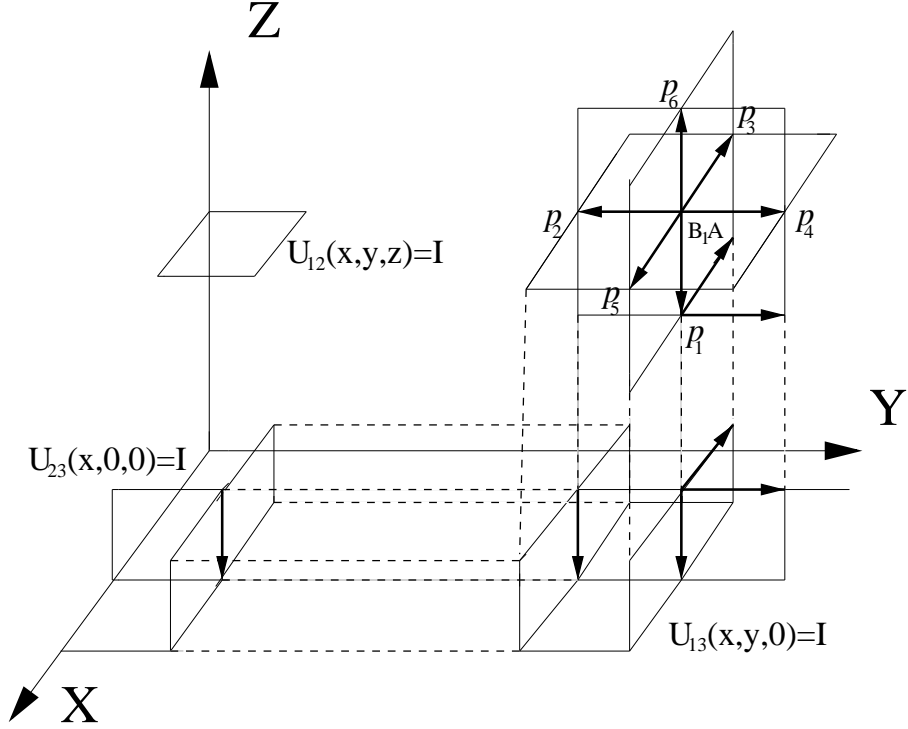}}
\caption{\label{delta} Graphical representation on the dual lattice
of the cube delta-function for the cube $B_1A$.}
\end{figure}

Now we can write down the plaquette representation for the $SU(N)$ gauge
model on the dual lattice
\begin{eqnarray}
Z_{\Lambda}(\beta , m_f, N_f) &=& \int \prod_{l}d V_{l}  
\exp \left [ \beta \sum_l H [V_l] + \sum_{f=1}^{N_f} {\rm  Tr} 
\ln A_f \left [ c(x),c(x^{\prime}); U_l[V_l] \right ] \right ] 
\nonumber   \\ 
&\times& \prod_{i=1}^4  \ \prod_{x(i)} \ J\left ( V_x^{(i)} \right ) \ ,
\label{dualpart}
\end{eqnarray}
\noindent
where $i=1,2,3,4$ denote four BCC sub-lattices, $\prod_{x(i)}$ runs over all
sites of such a BCC sub-lattice and $J(V_x^{(i)})$ is an $SU(N)$ delta-function
(\ref{func2}). To visualise calculations in the next sections we give here
full expressions for the matrices $V_x^{(i)}$. As example, we also give
the graphical representation of the Bianchi identity for the cube $B_1A$
in Fig.\ref{delta}. To write down the full expressions we choose a coordinate
system on the dual lattice as shown in Fig.\ref{delta}.
Let $\vec{x_i}=(x_i,y_i,z_i)$ be a site of the dual lattice which belongs
to the $i$-th BCC sub-lattice and introduce notations for dual links 
\begin{eqnarray}
l_1=(\vec{x_i};n_1) \ , \ l_2=(\vec{x_i};n_2) \ , \ l_3=(\vec{x_i};n_3) \ ,
\nonumber  \\
l_4=(\vec{x_i}-e_1;n_1) \ , \ l_5=(\vec{x_i}-e_2;n_2) \ ,
\ l_6=(\vec{x_i}-e_3;n_3) \ .
\label{dualln}
\end{eqnarray}
\noindent
We keep these notations for links attached to any site $x_i$.
Then, for the type of the connector for the cube $B_1A$ one has
\begin{equation}
V_x^{(1)} =  V_{l_5}^{\dag}V_{l_1}V_{l_6}^{\dag} \
C_{\vec{x}(1)} \ V_{l_2}V_{l_3}V_{l_4}^{\dag} \ C_{\vec{x}(1)}^{\dag} \ ,
\label{V1}
\end{equation}
\noindent
\begin{equation}
C_{\vec{x}(1)} = \prod_{k=z_i-1}^1 V_{n_2}(x_i,y_i-e_2,k)
\prod_{j=L-1}^{x_i}V_{n_3}(j,y_i,0)
\prod_{p=1}^{z_i-1}V_{n_1}(x_i-e_1,y_i,p) \ ,
\label{CB1A}
\end{equation}
\noindent
for the connector of the cube $B_2A$ 
\begin{equation}
V_x^{(2)} =  V_{l_4}^{\dag}V_{l_5}^{\dag}V_{l_6}^{\dag} \
C_{\vec{x}(2)} \ V_{l_1}V_{l_3}V_{l_2} \ C_{\vec{x}(2)}^{\dag} \ ,
\label{V2}
\end{equation}
\noindent
\begin{equation}
C_{\vec{x}(2)} = \prod_{k=z_i-1}^1 V_{n_1}(x_i-e_1,y_i,k)
\prod_{j=x_i}^{L-1}V_{n_3}(j,y_i,0)
\prod_{p=1}^{z_i-1}V_{n_2}^{\dag}(x_i,y_i,p) \ ,
\label{CB2A}
\end{equation}
\noindent
for the connector of the cube $B_3A$ 
\begin{equation}
V_x^{(3)} =  V_{l_2}V_{l_4}^{\dag}V_{l_6}^{\dag} \ C_{\vec{x}(3)} \
V_{l_5}^{\dag}V_{l_3}V_{l_1} \ C_{\vec{x}(3)}^{\dag} \ ,
\label{V3}
\end{equation}
\noindent
\begin{equation}
C_{\vec{x}(3)} = \prod_{k=z_i-1}^1 V_{n_2}^{\dag}(x_i,y_i,k)
\prod_{j=L-1}^{x_i+1}V_{n_3}^{\dag}(j,y_i,0)
\prod_{p=1}^{z_i-1}V_{n_1}^{\dag}(x_i,y_i,p) \ ,
\label{CB3A}
\end{equation}
\noindent
and for the connector of the cube $B_4A$ 
\begin{equation}
V_x^{(4)} =  V_{l_1}V_{l_2}V_{l_6}^{\dag} \ C_{\vec{x}(4)} \
V_{l_4}^{\dag}V_{l_3}V_{l_5}^{\dag} \ C_{\vec{x}(4)}^{\dag} \ ,
\label{V4}
\end{equation}
\noindent
\begin{equation}
C_{\vec{x}(4)} = \prod^{1}_{k=z_i-1}V_{n_1}^{\dag}(x_i,y_i,k)
\prod_{j=x_i+1}^{L-1}V_{n_3}^{\dag}(j,y_i,0)
\prod_{p=1}^{z_i-1}V_{n_2}(x_i,y_i-e_2,p) \ .
\label{CB4A}
\end{equation}
\noindent
Finally, the fermionic matrix in the gauge (\ref{func5}) can be presented 
on the dual lattice as 
\begin{eqnarray}
A_f(c(x),c(x^{\prime});U_l[V_l]) = M_f\delta_{c,c^{\prime}} +
\frac{1}{2}  \left [ \delta_{c+e_3,c^{\prime}}\xi_3(c) 
+ \delta_{c-e_3,c^{\prime}} \overline{\xi}_3(c^{\prime}) \right ] 
\nonumber   \\ 
+ \frac{1}{2} \sum_{n=1}^2  \left [ \delta_{c+e_n,c^{\prime}}\xi_n(c) 
U_n[c;V_l] + 
\delta_{c-e_n,c^{\prime}} \overline{\xi}_n(c^{\prime})
U_n^{\dagger}[c^{\prime};V_l] \right ] \ ,
\label{fermmatrplaq}
\end{eqnarray}
\noindent
where $c(x)$ is a cube dual to the site $x$. $U_n[c;V_l]$ is given in 
(\ref{linku1}), (\ref{linku2}) for $n=2$ and similar expressions can be 
easily written for $n=1$.  

The expressions for the connectors $C_{\vec{x}(i)}$ in (\ref{V1})-(\ref{CB4A})
are given for the free BC. 
For the Dirichlet BC one should omit the second product $\prod_jV_{n_3}(j)$ 
from formulae (\ref{CB1A}), (\ref{CB2A}), (\ref{CB3A}) and (\ref{CB4A}). 
Extension of this representation to the lattice with periodic BC in all 
directions is, at least in principle very simple. Here we give a qualitative 
explanation only, the details will be given elsewhere. Instead of the maximal axial 
gauge which is not compatible with the periodic BC one can fix the following gauge 
\begin{equation}
U_3(x,y,z) = I \ , \ z\in [1, L-1] \ ; \ 
U_3(x,y,0) = W(\vec{x})=\prod_{z=0}^{L-1} U_3(x,y,z) \ ,
\label{plgauge}
\end{equation}
where $W$ is the Polyakov line. All the steps performed above are unchanged 
except for the integration over space-like links in the time slice $z=0$. 
In this slice the expressions for connectors are modified: they include the values 
of the Polyakov lines. Therefore, one can express the final result in terms 
of plaquettes and these Polyakov lines (\ref{plgauge}).

\subsection{Observables on the dual lattice}

In this subsection we derive on the dual lattice a representation for 
the t' Hooft line and the Wilson loop. 
        
We consider first a t' Hooft line which in $3D$ consists of a string $S_{xy}$ 
of dual links between the sites $x$ and $y$ on the dual lattice. For simplicity 
let us restrict ourselves to pure gauge theory with the Wilson action. 
With the set $T$ of plaquettes dual to the string $S_{xy}$ we define the action 
\begin{equation}
S_z(U;T) = \beta\sum_{p\notin T}{\rm Re Tr}U_p + 
\beta\sum_{p\in T}{\rm Re Tr}zU_p 
\label{Taction}
\end{equation}
\noindent
and the partition function
\begin{equation}
Z(z;S_{xy}) = \int DU \ \exp \{ \beta S_z(U;T) \} \ ,
\label{TPF}
\end{equation}
\noindent
where $z$ is center element of $SU(N)$. We can introduce the disorder 
operator as 
\begin{equation}
\langle D_z(S_{xy}) \rangle = \frac{Z(z;S_{xy})}{Z(1;S_{xy})} \ .
\label{disop}
\end{equation}
\noindent	
On the dual lattice the partition function $Z(z,S_{xy})$ of pure gauge theory 
can be written in the form of Eq.(\ref{dualpart}) with the action 
\begin{equation}
S_z(V;S_{xy}) = \beta\sum_{l\notin S_{xy}}{\rm Re Tr}V_l + 
\beta\sum_{l\in S_{xy}}{\rm Re Tr}zV_l \ . 
\label{Sxyaction}
\end{equation}
\noindent	
Suppose that the points $x$ and $y$ belong to the BCC sub-lattice with the $i=1$ 
type of connector. By a change of variables $V_l\to z^{*}V_l$ for $l\in S_{xy}$ 
we get the following representation for the partition function $Z(z;S_{xy})$ 
\begin{eqnarray}
Z(z;S_{xy}) = \int \prod_{l}d V_{l}  \exp \left [ \beta \sum_{l} 
{\rm Re \ Tr} V_{l} \right ] \ \prod_{i=1}^3  \ \prod_{x(i)} \ 
J\left ( V_x^{(i)} \right )  
\nonumber  \\ 
\times \left [ \prod_{x(1)\ne x,y} \ J\left ( V_x^{(1)} \right ) \right ] \ 
J\left ( zV_x^{(1)} \right ) J\left ( zV_y^{(1)} \right ) \ ,
\label{SxyPF}
\end{eqnarray}
\noindent	
since $z$ cancels from all Bianchi constraints but those for the endpoints. 
Therefore, one can say that the Bianchi constraint is violated at these 
endpoints. For example, for the $SU(2)$ gauge group $z=-1$ and $J(zV)$ has the form 
\begin{equation}
\label{BIviol}
J(zV) =  \delta (V+I) = \sum_r (-1)^{2r}d_r \chi_r \left ( V \right ) \ .
\end{equation}
\noindent
This delta-function constrains the product of non-abelian matrices to $-I$.

Now we give a dual expression for a Wilson loop of size $R\times T$ 
in some representation $j$. Suppose, for simplicity that the loop contour 
lies in the $y-z$ plane and $R$, $T$ are even. Generelizing 
the consideration of the previous subsection it is straightforward to show 
that such a Wilson loop has the following dual form 
\begin{eqnarray}
W_j(C) =  \langle \ {\rm Tr }_j \ \prod_{n=0}^{R/2-1} 
\left( \prod_{z_1=z-1}^0 V_1(x,y+2n,z_1) 
\prod_{z_2=0}^{z-1} V_1 (x,y+2n+1,z_2) \right ) \nonumber  \\
\prod_{n=R/2-1}^{0} \left ( \prod_{z_1=0}^{z+T-1} V_1^{\dag}(x,y+2n+1,z_1)
\prod_{z_2=z+T-1}^0 V_1^{\dag}(x,y+2n,z_2) \right ) \  \rangle \ .
\label{WLdual}
\end{eqnarray}
As one can see, the connectors are not cancelled in the general from the Wilson loop.
One can simplify this expression taking one side of the loop in the plane $z=0$.
Then the Wilson loop is an ordered product of dual links which all lay inside the  
area spanned by the loop $C$, i.e. the first line in the last formula is unity.

\section{Low-temperature expansion}

This section is devoted to the construction of the low-temperature
expansion of gauge models in the plaquette formulation. In doing this
we follow the strategy developed for the two-dimensional principal chiral models
in \cite{2dsunlink}. In that paper a low-temperature expansion was constructed
for $SU(N)$ spin models in the link formulation. This formulation appears
to be closely related to the plaquette formulation of $3D$ gauge models. For example,
as was shown in \cite{plrepr} and \cite{linkrepr} in both cases the strong coupling
expansion is the expansion towards restoration of the Bianchi identity.
As will be seen below, the weak-coupling expansion in both models also bears many
common features. In the abelian case they turn out to be practically identical,
while the only (but essential) difference in the non-abelian gauge models is 
the appearance of connectors.

A lattice PT in the maximal axial gauge was constructed
for abelian and non-abelian models in the standard Wilson formulation
by F.~M\"uller and W.~R\"uhl in \cite{pttemp}.
In this gauge the PT for non-abelian models displays serious infrared
divergences in separate terms of the expansion starting from
${\cal O}(\beta^{-2})$ order for the Wilson loop and from
${\cal O}(\beta^{-1})$ for the free energy. Of course, it is expected that
all divergences must cancel in all gauge invariant quantities.
In \cite{pttemp} the authors worked out a special procedure of how to deal with
such divergences. The essential ingredients of the procedure of \cite{pttemp}
are the following: 1) fixing the Dirichlet BC in one direction and the periodic
ones in other directions; 2) a Wilson loop under consideration is placed at distance
$R$ from the boundary and the limit $R\to\infty$ is imposed to restore
time translation invariance; 3) divergent Green functions must be properly
regularised, though there is no a priori preference to any regularization.
With this procedure the first two coefficients of the expansion of Wilson
loop expectation values in $SU(2)$ LGT have been computed. These coefficients 
appear to be infrared finite in all $D\geq 3$ and are proportional to 
the perimeter of the loop in $4D$ and to $R\ln T + T \ln R$ in $3D$ 
(for a rectangular loop $R\times T$).

We are not aware
of any proof of the infrared finiteness at higher orders of the expansion.
That there could be a problem with the infrared behaviour was shown however
in \cite{seiler}. Namely, in the addition to the Dirichlet BC on the boundary,
one link was fixed in the middle of the lattice and it was shown that the TL
value of the expectation value of the plaquette to which the fixed link belong
differs from that obtained in \cite{pttemp}.
The underlying reason of such infrared behaviour lies, of course in the fact
that the conventional PT is done around the vacuum $U_{n}(x)=I$ which is the true
ground state only at fixed $L$, even in the maximal axial gauge. When the volume
grows the fluctuations of the link matrices become larger and larger and
may cause an infrared unstable behaviour. In particular, the integration regions
over fluctuations are usually extended to infinity. But only for fixed $L$ one
can really prove that this introduces exponentially small corrections.
There is no tool for proving that they remain exponentially small also in
the large volume limit. On the contrary, bounds on the large plaquette
fluctuations (\ref{chestplaq}) holds uniformly in the volume implying
that large plaquette fluctuations remain exponentially small also in the TL.
This is the first achievement of the low-temperature expansion in the
plaquette formulation.
Constructing the low-temperature expansion in the plaquette formulation we aim
not only to develop a new technics of calculation of the asymptotic expansions
in LGT but also to get a deeper insight into this infrared problem and
as a consequence into the problem of the uniformity of the low-temperature 
expansion. We hope that this will help
in the investigation of the general properties of asymptotic expansions
of non-abelian gauge models. As will be seen below, all Green functions
appearing in the plaquette formulation are infrared finite. The source
of trouble are the connectors of the non-abelian Bianchi identity. The low-temperature
expansion in the plaquette formulation starts from the abelian Bianchi identity
(zero order partition function) for every cube and goes towards gradual restoration
of the full identity with every order of the expansion. Thus, the generating
functional contains an abelianized form of the identity without connectors.
High-order terms include expansion of connectors (of course, together with
other contributions which are however infrared finite) which can lead to
the infrared problem. Let us also mention that in this respect there is a big
difference between gauge models and $2D$ $SU(N)$ spin models.
In the latter the low-temperature expansion also goes towards restoration
of the full non-abelian Bianchi identity \cite{2dsunlink} and the form of
the generating functional is formally the same. It includes invariant
link Green functions and the standard Green function.  In $2D$ the latter is not 
infrared finite and is the main source of the trouble. As will be seen below, 
in $3D$ gauge models all Green functions entering the generating functional are 
infrared finite. The source of the infrared problem in gauge models are 
the connectors of the non-abelian Bianchi identities.

\subsection{$U(1)$ LGT}

For a simple introduction to our method we first consider
the abelian $U(1)$ LGT without fermions where the expansion can be done 
in a straightforward manner. 
Due to a cancellation of the connectors from Bianchi identities for 
abelian models the plaquette formulation for the $U(1)$ model on the dual lattice 
with the Wilson action reads
\begin{equation}
Z_{U(1)}(\beta ) = \int_{-\pi}^{\pi} \prod_l \frac{d\omega_l}{2\pi}
\exp \left [\beta \sum_l\cos\omega_l \right ]\prod_xJ_x \ ,
\label{PFLu1}
\end{equation}
\noindent
where the Jacobian is given by the periodic delta-function
\begin{eqnarray}
\label{PFLu1J}
J_x &=& \sum_{r=-\infty}^{\infty} e^{ir\omega_x} \ , \\
\omega_x &=& \omega_{n_1}(x)+\omega_{n_2}(x) +\omega_{n_3}(x) -
\omega_{n_1}(x-e_1)-\omega_{n_2}(x-e_2)-\omega_{n_3}(x-e_3) \ . \nonumber
\end{eqnarray}
\noindent
Dual links are defined by the point $x$ and the positive
direction $n$, see Eq.(\ref{dualln}). Strictly speaking, if the original
link gauge matrices satisfy free boundary conditions, and so do the plaquette
matrices then the dual links $\omega_l$ and representations $r_x$ must obey
zero Dirichlet BC. All general expansions given below are valid for any type of BC,
the dependence on the BC enters only the Green functions defined below.
In $3D$ the difference between Green functions for different types of BC is
of the order ${\cal O}(L^{-2})$ for large $L$. Since we are interested in the TL
behaviour only, we can consider dual models without any reference to the original
link representation and introduce any type of BC. Both in abelian and in non-abelian
cases we work with periodic BC. 
We remind at this point that expressions obtained below
on a finite lattice do not correspond to the standard link formulation since
the latter would include contributions from nontrivial Polyakov loops
on a periodic finite lattice. However, in the TL both models must coincide and 
convergence to the TL is very fast, like ${\cal O}(L^{-2})$. 
If one wants to recover the exact correspondence between our dual representation and 
the link representation on the finite lattice with free BC, the only change is 
that one should take the corresponding Green functions entering the generating 
functionals.

The construction of the PT proceeds as follows. 
The first step is a standard one, i.e. we re-scale
$\omega\to \omega/\sqrt{\beta}$ and expand the Boltzmann factor
in powers of $1/\beta$ as
\begin{equation}
\exp \left[ \beta\cos\frac{\omega_l}{\sqrt{\beta}} \right] = \exp
\left[ \beta -\frac{1}{2}(\omega_l)^2 \right]
\left[ 1 + \sum_{k=1}^{\infty} (\beta)^{-k}A_k(\omega_l^2) \right] \ ,
\label{cos}
\end{equation}
\noindent
where $A_k$ are known coefficients (see, e.g. \cite{2dsunlink}).
Introducing sources $h_l$ for the link field, we find
\begin{equation}
Z_{U(1)}(\beta >> 1) = e^{3\beta L^3 - L^3\ln \beta}
\prod_{l} \left[ 1+\sum_{k=1}^{\infty}\frac{1}{\beta^k}
A_k\left( \frac{\partial^2}{\partial h_l^2} \right) \right] M_{U(1)}(h_l) \ ,
\label{asu1Gll}
\end{equation}
\noindent
where $M_{U(1)}(h_l)$ is a zero-order generating functional. 
Using the Poisson summation formula $M_{U(1)}(h_l)$ can be represented as
\begin{eqnarray}
M_{U(1)}(h_l) = \sum_{m_x=-\infty}^{\infty}\int_{-\infty}^{\infty}
\prod_xdr_x \int_{-\pi\sqrt{\beta}}^{\pi\sqrt{\beta}} 
\prod_l \frac{d\omega_l}{2\pi} 
\nonumber  \\
\times \exp \left[-\frac{1}{2}\sum_l \omega_l^2
+ i\sum_l\omega_l (r_x-r_{x+e_n}) + 2\pi i \sqrt{\beta}\sum_xr_xm_x
+\sum_l\omega_l h_l  \right] \ .
\label{Gfunu1}
\end{eqnarray}
\noindent
In addition to the above perturbation one has to extend the integration region 
over dual link angles $\omega_l$ to infinity. 
Clearly, all the corrections from this perturbation go down
exponentially with $\beta$ (it is obvious from the expression for the generating
functional, Eq.(\ref{Gfunu1})).
Calculating all the integrals in (\ref{Gfunu1})
we find up to a constant (sum over all repeating indices is understood)
\begin{equation}
M_{U(1)}(h_l)=e^{\frac{1}{4}h_lG_{ll^{\prime}}h_{l^{\prime}}}
\sum_{m_x}e^{-\pi^2\beta m_xG_{x,x^{\prime}}m_{x^{\prime}}+
\pi\sqrt{\beta}h_lD_l(x^{\prime})m_{x^{\prime}}} \ ,
\label{MU1}
\end{equation}
\noindent
where we have introduced the link Green functions $G_{ll^{\prime}}$ and
$D_l(x)$ (see Appendix A). $G_{x,x^{\prime}}$ is the Green function of 
free massless scalar field in $3D$. Up to the first exponent, the generating 
functional coincides with the dual partition function of the $U(1)$ Villain 
model in the Coulomb gas representation. From here it follows: Only the 
configuration $m_x=0$ for all $x$ contributes to the asymptotic expansion and 
non-trivial monopole configurations are exponentially suppressed.
We thus get
\begin{equation}
M_{U(1)}(h_l) = \exp
\left[ \frac{1}{4} \sum_{l,l^{\prime}}h_lG_{ll^{\prime}}h_{l^{\prime}} \right] \
+ {\cal O}(e^{-\beta}) \ .
\label{Gfunu2}
\end{equation}
\noindent
Substitution of the last equation into (\ref{asu1Gll}) generates 
a weak-coupling expansion in inverse powers of $\beta$. 

The easiest way to construct the corresponding expansion for the
Wilson loop is the following. Let $S_{xy}^d$ be some surface dual to the surface 
$S_{xy}$ which is bounded by the loop $C$ and consists of links dual to 
plaquettes of the original lattice. Let $b$ denote links from $S_{xy}^d$. 
Then, the Wilson loop in representation $J$ is defined as
\begin{equation}
W(C) = \langle \prod_{b\in S^d_{xy}} 
\exp \left [ \frac{i}{\sqrt{\beta}} J \omega_b \right ] \rangle \ .
\label{U1Wl}
\end{equation}
\noindent
The last formula suggests that for the computation of the Wilson loop 
it is sufficient to make the shift 
$h_l\to h_l+(i/\sqrt{\beta})J\sum_b\delta_{l,b}$
in the expression for the generating functional (\ref{Gfunu2}).
We find
\begin{eqnarray}
W(C) = \exp \left [- \frac{J^2}{4\beta}
\sum_{b,b^{\prime}\in S^d_{xy}} 
G_{bb^{\prime}}\right ] 
Z^{-1}_{U(1)} \prod_{l} \left[ 1+\sum_{k=1}^{\infty}\frac{1}{\beta^k}
A_k\left( \frac{\partial^2}{\partial h_l^2} \right) \right] \nonumber  \\
\exp \left[ \frac{1}{4} \sum_{l,l^{\prime}}h_lG_{ll^{\prime}}h_{l^{\prime}} 
+ \frac{i}{2\sqrt{\beta}} J h_l\sum_{b}G_{lb} \right ] \ .
\label{U1WLlt}
\end{eqnarray}
\noindent
It is now straightforward to calculate all connected pieces contributing
to the Wilson loop at a given order of $1/\beta$. For the first coefficients we get
\begin{equation}
W(C) = \exp \left [ - \frac{J^2}{2\beta} 
\left ( 1+\frac{1}{4\beta} \right ) P(C) + 
{\cal O} \left ( \beta^{-3} \right ) \right ] \ ,
\label{WLexp}
\end{equation}
\noindent
where 
\begin{equation}
2P(C) = \sum_{b,b^{\prime}\in S^d_{xy}} G_{bb^{\prime}} \ .
\label{W2}
\end{equation}
\noindent
Using some properties of $G_{ll^{\prime}}$ described in the Appendix A 
it is easy to prove that this result coincides with the result given  
in \cite{pttemp} for the $U(1)$ model and for $J=1$. 
The asymptotic behaviour of $P(C)$ is given by Eq. (\ref{PCTL}).

\subsection{$SU(2)$ LGT}

In this subsection we derive the low-temperature expansion for non-abelian
models. Again, the general procedure is precisely the same as in $2D$
non-abelian spin models, therefore for some technical details
we refer the reader to paper \cite{2dsunlink}.
First, we describe the general procedure for pure $SU(2)$ LGT and then
give a simple generalisation for all $SU(N)$ models and include dynamical 
fermions. For simplicity we work in what follows with the Wilson action, 
i.e. $H[V_l]={\rm Re \ Tr} V_l$. 

We want to expand the partition and correlation functions
into an asymptotic series whose coefficients $\langle B_k \rangle_{G}$ 
are calculated in some ensemble $G$ with a Gaussian measure. 
For the partition function we expect up to a constant factor an expression 
of the form
\begin{equation}
Z = 1+\sum_{k=1}^{\infty}\frac{1}{\beta^k}
\langle B_k \rangle_{G} \ .
\label{AppA2}
\end{equation}
Let us consider the standard parameterisation for the $SU(2)$ link matrix
\begin{equation}
V_l = \exp [i\sigma^k\omega_k(l)] \ ,
\label{su2mtr}
\end{equation}
\noindent
where $\sigma^k, k=1,2,3$ are Pauli matrices and introduce
\begin{equation}
W_l = \left[ \sum_k\omega^2_k(l) \right]^{1/2} \ \ , \
W_x = \left[ \sum_k\omega^2_k(x) \right]^{1/2} ,
\label{Wp}
\end{equation}
\noindent
where $\omega_k(x)$ is a site angle defined as
\begin{equation}
V_x =  \exp \left [ i\sigma^k\omega_k(x) \right ] =
\left ( \prod_{p \in A} V_p \right ) \ C
\left ( \prod_{p \in B} V_p \right ) \ C^{\dag} \ .
\label{plangle}
\end{equation}
\noindent
Here, the site $x$ is dual to the cube $c$ with vertices $A$ and $B$.
$\omega_k(x)$ has the following expansion in powers of link angles
\begin{equation}
\omega_k(x) =  \omega^{(0)}_k(x) +
\omega^{(1)}_k(x) + \omega^{(2)}_k(x) + \cdots \ .
\label{AppA3}
\end{equation}
\noindent
On the dual lattice (we keep notations (\ref{dualln}) for dual links)
the first coefficient can be written down as
\begin{equation}
\omega^{(0)}_k(x) = \omega_k(l_4) + \omega_k(l_5) + \omega_k(l_6) -
\omega_k(l_1) - \omega_k(l_2) - \omega_k(l_3) \ .
\label{wkx0}
\end{equation}
\noindent
$\omega^{(i)}_k(x)$ for $i\geq 1$ can be computed by making repeated use
of the Campbell-Baker-Hausdorff formula.
Then, the partition function (\ref{dualpart}) can be exactly rewritten
to the following form (see Appendix A in the e-print version of \cite{2dsunlink})
\begin{eqnarray}
Z_{SU(2)} = \int \prod_l \left[ \frac{\sin^2W_l}{W^2_l}\prod_kd\omega_k(l) \right]
\exp \left[ 2\beta\sum_l\cos W_l \right] \prod_x \frac{W_x}{\sin W_x}
\nonumber     \\
\prod_x \sum_{m(x)=-\infty}^{\infty}\int\prod_kd\alpha_k(x)
\exp \left [ -i\sum_k\alpha_k(x)\omega_k(x) + 2\pi im(x)\alpha (x) \right ] \ ,
\label{PFwk}
\end{eqnarray}
\noindent
where $\alpha (x)=(\sum_k\alpha^2_k(x))^{1/2}$.
For the derivation of this representation we have used the Poisson
resummation formula.
In order to perform the weak coupling expansion we make the substitution
\begin{equation}
\omega_k(l)\to (2\beta)^{-1/2}\omega_k(l) \ , \
\alpha_k(x)\to (2\beta)^{1/2}\alpha_k(x)
\label{subst}
\end{equation}
\noindent
and then expand the integrand of (\ref{PFwk})
in powers of fluctuations of the link fields.
We introduce now the external sources $h_k(l)$ coupled to the link field
$\omega_k(l)$ and $s_k(x)$ coupled to the auxiliary field $\alpha_k(x)$
and adopt the definitions
\begin{equation}
\omega_k(l) \to \frac{\partial}{\partial h_k(l)} \ , \
\alpha_k(x) \to \frac{\partial}{\partial s_k(x)} \ .
\label{deriv}
\end{equation}
\noindent
With this convention we get the following expansion
for the partition function
(\ref{PFwk})
\begin{equation}
Z = \left [ 1+\sum_{k=1}^{\infty}\frac{1}{\beta^k}
B_k\left (\partial_h,\partial_s \right ) \right ] M(h,s) \ ,
\label{AppA1}
\end{equation}
where the operators $B_k$ are defined through
\begin{eqnarray}
1 &+& \sum_{k=1}^{\infty}\frac{1}{\beta^k}
B_k\left (\partial_h,\partial_s \right ) =  \prod_x
\left[ 1 + \sum_{q=1}^{\infty}\frac{(-i)^q}{q!} \left( \sum_k\alpha_k(x)
\sum_{n=1}^{\infty}
\frac{\omega^{(n)}_k(x)}{(2\beta )^{n/2}} \right)^q  \right] \nonumber \\
&\times& \prod_l\left[ \left( 1 + \sum_{k=1}^{\infty} \frac{1}{(2\beta )^k}
\sum_{l_1,..,l_k}^{\prime}\frac{a_1^{l_1}...a_k^{l_k}}{l_1!...l_k!} \right)
\left( 1 + \sum_{k=1}^{\infty}\frac{(-1)^k}{(2\beta )^k}C_kW_l^{2k}
\right) \right]   \ .
\label{PFwkexp}
\end{eqnarray}
\noindent
Here we have denoted 
\begin{equation}
a_k = (-1)^{k+1}\frac{W_l^{2(k+1)}}{(2k+2)!} \ , \
C_k = \sum_{n=0}^k \frac{1}{(2n+1)!(2k-2n+1)!} \ .
\label{a_kW}
\end{equation}
\noindent
The first brackets on the rhs in Eq.(\ref{PFwkexp}) represent the expansion
of the Jacobian. The first and the second brackets in the second line
come from the expansion of the action and of the invariant measure,
correspondingly. We have omitted the expansion of the term $W_x/\sin W_x$
since it does not contribute to the asymptotic expansion (due to the
constraint $W_x=0$) but only to exponentially small corrections. 
As usual, one has to put $h_k=s_k=0$ after taking all the derivatives.
Note, that in writing down these general expansions we do not distinguish
between four BCC sub-lattices since the generating functional (see just below)
has a unique form for all lattice. Thus, the product over $x$ in Eq.(\ref{PFwkexp})
goes over all sites of the dual lattice. The difference between BCC sub-lattices
appears only in the general expressions for coefficients $\omega^{(i)}_k(x)$
for $i\geq 1$.

The generating functional $M(h,s)$ is given by
\begin{equation*}
M(h,s) = \int_{-\infty}^{\infty} \prod_{x,k} d\alpha_k(x)
\int_{-\infty}^{\infty} \prod_{l,k} d\omega_k(l)
\exp [ -\frac{1}{2}\omega^2_k(l)
-i\omega_k(l)[\alpha_k(x+e_n) - \alpha_k(x) ] ]
\end{equation*}
\begin{equation}
\times \ \sum_{m(x)=-\infty}^{\infty}
\exp \left[ 2\pi i\sqrt{2\beta}\sum_xm(x)\alpha (x) +
\sum_{l,k}\omega_k(l)h_k(l) + \sum_{x,k}\alpha_k(x)s_k(x) \right] \ ,
\label{GF1}
\end{equation}
\noindent
where we extended the integration region for the link angles $\omega_k(l)$ 
to the real axes. As in the abelian case, only the configuration with $m_x=0$ 
for all $x$ contributes to the asymptotic expansion, all others are
exponentially suppressed. Calculating all Gaussian integrals
in (\ref{GF1}) we come to
\begin{equation}
M(h,s) = \exp
\left[ \frac{1}{4}s_k(x)G_{x,x^{\prime}}s_k(x^{\prime}) +
\frac{i}{2}s_k(x)D_l(x)h_k(l) +
\frac{1}{4}h_k(l)G_{ll^{\prime}}h_k(l^{\prime}) \right] \ .
\label{GFfin}
\end{equation}
\noindent
From the last expression one can deduce the following simple rules
\begin{equation}
\langle \omega_k(l)\omega_n(l^{\prime}) \rangle =
\frac{\delta_{kn}}{2}G_{ll^{\prime}} \ , \
\langle \alpha_k(x)\alpha_n(x^{\prime}) \rangle =
\frac{\delta_{kn}}{2}G_{xx^{\prime}} \ ,
- i\langle \omega_k(l)\alpha_n(x^{\prime}) \rangle =
\frac{\delta_{kn}}{2}D_l(x^{\prime}) \ .
\label{Frules}
\end{equation}
\noindent
The expansion (\ref{AppA1}), the representation (\ref{GFfin}) for the
generating functional and the rules (\ref{Frules})
are the main formulae of this section which allow
to calculate the weak coupling expansion of both the free energy and any
fixed-distance observable. The extension of this expansion for the Wilson
loop is straightforward. Since all gauge invariant
quantities depend only on the dual link variables but not on auxiliary
fields one can make their direct expansion in powers of link angles.

\subsection{$SU(N)$ LGT and dynamical fermions}

A generalisation for an arbitrary $SU(N)$ model can be done again precisely like
in $2D$ spin models. For completeness, we repeat the arguments
of \cite{2dsunlink} below. It is seen from the procedure outlined above that
the low-temperature expansion arises only from the ``vacuum'' sector with
$m_x=0$ for all $x$. It follows from Eq.(\ref{PFwk}) that in this sector
the $SU(N)$ delta-function reduces to the Dirac delta-function so that
the partition function becomes
\begin{equation}
Z(\beta ) = \int \prod_l dV_l
\exp \left[ \beta \sum_l {\mbox {Re Tr}} V_l \right]
\int_{-\infty}^{\infty}\prod_{x}\prod_{k=1}^{N^2-1} d\alpha_k(x)
\exp \left [ -i\alpha_k(x)\omega_k(x) \right ] \ .
\label{lPFDLB}
\end{equation}
\noindent
After this simple observation the expansion itself is done precisely 
like for $SU(2)$.

To include fermions in the above expansion one can proceed in 
a standard way. Namely, we expand gauge matrices entering the fermionic 
matrix (\ref{fermmatrplaq}) as 
\begin{equation}
U_n(c) = \exp [it_k\omega_k(c,n)] = \sum_{m=0}\frac{i^m}{m!} 
[t_k\omega_k(c,n)]^m \ .
\label{unferm}
\end{equation}
Here $\omega_k(c,n)=\omega_k(p)$ can be calculated from equations 
(\ref{linku1}) and (\ref{linku2}) with the help of the Campbell-Baker-Hausdorf 
formula. We introduce now the following matrices 
\begin{equation}
\hat{G}_f^{-1}(c(x),c(x^{\prime})) = M_f\delta_{c,c^{\prime}} +
\frac{1}{2} \sum_{n=1}^3 
\left [ \delta_{c+e_n,c^{\prime}}\xi_n(c) 
+ \delta_{c-e_n,c^{\prime}} \overline{\xi}_n(c^{\prime}) \right ] 
\label{fermmatrfree}
\end{equation}
\noindent
and 
\begin{eqnarray}
F(c(x),c(x^{\prime});U_l[V_l]) = \sum_{m=1}\frac{i^m}{m!} 
\sum_{k_1...k_m}F^{(m)}(c(x),c(x^{\prime}))t_{k_1}...t_{k_m} \ ,  
\label{fermmatrinteract}
\end{eqnarray}
\noindent
where
\begin{eqnarray}
F^{(m)}(c(x),c(x^{\prime})) &=&  
\frac{1}{2} \sum_{n=1}^2   [ \delta_{c+e_n,c^{\prime}}\xi_n(c) 
\omega_{k_1}(c,n)...\omega_{k_m}(c,n) 
\nonumber   \\
&+& (-1)^m  
\delta_{c-e_n,c^{\prime}} \overline{\xi}_n(c^{\prime})
\omega_{k_1}(c^{\prime},n)...\omega_{k_m}(c^{\prime},n) ] \ .
\label{Fcc}
\end{eqnarray}
\noindent
Then, the fermionic part of the partition function (\ref{dualpart}) is 
expanded as 
\begin{eqnarray}
\exp \left [ \sum_{f=1}^{N_f} {\rm  Tr} 
\ln A_f \left [ c(x),c(x^{\prime}); U_l[V_l] \right ] \right ] = 
\prod_{f=1}^{N_f}{\rm Det}[\hat{G}_f^{-1}] 
\nonumber  \\  
\times \exp \left [ \sum_{f=1}^{N_f}\sum_{s=1}^{\infty}\frac{(-1)^s}{s} 
{\rm Tr} (\hat{G}_f F)^s  \right ] \ .
\label{fermptexp}
\end{eqnarray}
With the substitution (\ref{subst}) and expressing the plaquette angles 
$\omega_k(c,n)$ in terms of $\omega_k(l)$ the last formula can be expanded 
further in powers of fluctuations of dual link variables. This procedure 
obviously does not change the generating functional therefore all 
expectation values can be obtained again by differentiating the functional 
(\ref{GFfin}).  

In order to show how our expansion works in practice we have computed
the simplest possible quantity, namely the first order coefficient in the 
expansion of the free energy. All details of these calculations for $SU(N)$ 
model are given in the Appendix C. 

\subsection{Some remarks on the low-temperature expansion}

We close this section with brief comments on general properties of  
the weak coupling expansion in non-abelian models.
Consider the zero order partition function. As follows from
Eqs.(\ref{Gfunu1}) and (\ref{lPFDLB}) for the $SU(N)$ group it has the form
\begin{equation}
Z_G = \int_{-\infty}^{\infty} \prod_{l,k} d\omega_k(l)
\exp \left[ -\frac{1}{2} \sum_{l,k}\omega_k^2(l)  \right]
\prod_{x,k}\delta \left [ \omega_k^{(0)}(x) \right ] \ , \
k = 1, \cdots ,N^2-1 \ .
\label{ZG}
\end{equation}
\noindent
The integrations are performed over the real axes though originally they are 
restricted to a compact domain, e.g. for $SU(2)$ to (after change of variables 
(\ref{subst})) $\sum_k\omega_k^2(l)\leq \pi^2\beta$. From here it follows that the 
probability for $K$ links to have large fluctuations like
$\omega_k(l)\sim {\cal O}(\sqrt{\beta})$ is suppressed as
${\cal O}(e^{-{\rm const}K\pi^2\beta})$. This is of course
only a rough estimate, nevertheless, qualitatively
it is true and shows that large fluctuations are under good control since
they are exponentially suppressed with $\beta$, and this obviously remains
true in the TL. As we have already stressed, this property is an essential
achievement of the expansion in the plaquette representation. No such
control exists for large fluctuations of the link gauge matrices in the
large volume limit.

The next observation which can be deduced examining the zero order partition
function is that the low-temperature expansion is an expansion which starts
from the abelian Bianchi identity for each component of the non-abelian field
and goes towards restoration of the full non-abelian identity (compare with 
the strong coupling expansion, \cite{plrepr}).

It follows from the definition of the link Green function (\ref{Gll1}) that
\begin{equation}
\mid G_{ll^{\prime}} \mid \leq 4/3 \ .
\label{bound}
\end{equation}
\noindent
Therefore, with respect to the Gaussian measure in Eq.(\ref{Gfunu1})
the fluctuations of the link variables are bounded by
\begin{equation}
\mid \langle \omega_l\omega_{l^{\prime}} \rangle_G \mid =
\mid \frac{1}{2}G_{ll^{\prime}}\mid \leq \frac{2}{3} \ ,
\label{linkfluc}
\end{equation}
\noindent
due to the bound (\ref{bound}). One sees from the last formula
that the interaction between links (=original plaquettes) strongly decreases
with the distance. Taking the asymptotic expansion for the Green functions
(\ref{Gllasymp}) we find 
\begin{equation}
\mid \langle \omega_l\omega_{l^{\prime}} \rangle_G \mid =
{\cal O}(R^{-3}) \ .
\label{linkcorr}
\end{equation}
\noindent
This property justifies the low-temperature expansion in
powers of fluctuations of plaquette variables, while
there is no such justification for the expansion in terms
of the original link variables fluctuations which are not
bounded when $L\to\infty$.

The zero order partition function (\ref{ZG}) is equivalent to the $(N^2-1)$-component
free scalar field $\alpha_k(x)$. In our expansion it enters as auxiliary field.
As is seen from the generating functional (\ref{GFfin}) it depends only on
infrared finite Green functions. Link Green functions are infrared finite by
construction (see Appendix A), while the Green function for the free scalar
field is finite in any dimension $D\geq 3$. For $D=3$ one has in the TL
\begin{equation}
\mid G_{xx^{\prime}} \mid \leq G_0 = 0.5634 \ .
\label{boundGx}
\end{equation}
\noindent
Nevertheless this shows that while correlations of dual links and correlations
of links with auxiliary fields decrease with distance as is seen from
(\ref{Frules}) this is not the case for the correlations among auxiliary fields.
It follows from (\ref{Frules}) and the last bound that this correlation stays
constant however large the distances between fields are. Just this property together
with certain contributions from connectors creates an infrared problem in
non-abelian models. To be precise the problem appears in the following manner. 
Connectors generate contributions of the form $\sum_{l\in C}\omega_k(l)$ and 
of high order to all $\omega_k^{(i)}(x)$, see Eqs.(\ref{conanglexp}), 
(\ref{wkx1con})-(\ref{wkx2con}). And though all $\omega_k(l)$ behave like 
${\cal O} (1/\sqrt{\beta})$ it is not obvious if it remains true for the sums 
of the plaquette angles along connectors. Since in the TL practically all 
connectors become infinitely long, this raises the question whether 
the property $\sum_{l\in C}\omega_k(l)\sim {\cal O} (1/\sqrt{\beta})$ holds 
in the limit $L\to\infty$?

\section{Conclusions}

In this article we proposed a plaquette formulation of non-abelian lattice gauge 
theories. Our approach to such formulations is summarised in section 2.1. 
We have also included dynamical fermions in our construction.  
The main formula of section 2, Eq.(\ref{dualpart}), gives a plaquette 
formulation for gauge models with dynamical fermions.
As an application of our formulation we have developed a weak-coupling 
expansion which can be used for a perturbative evaluation of both the free 
energy and gauge invariant quantities like the Wilson loop. We believe that 
this work can be useful in at least two aspects. The first concerns the problem 
of the uniformity of the perturbative expansion in non-abelian models and is 
described in section 3. The second aspect concerns the perturbative 
expansion of lattice models with actions different from the Wilson action. 
Practically, all standard actions discussed in the literature have a very simple 
form in the plaquette formulation. The hardest part in the perturbative expansion 
is to treat contributions from the Jacobian. Since the Jacobian represents 
Bianchi constraints on the plaquette matrices and is the same whatever original 
action is taken it is sufficient to compute contributions from the Jacobian 
only one time and use the result for all actions. For example, the expressions for 
the coefficients $C_{meas}$, $C_{J1}$ and $C_{J2}$ in (\ref{frensum}) given by 
(\ref{c1meas}), (\ref{CJ1res}) and (\ref{CJ2ref}), respectively must be 
the same for all lattice actions. The expression for $C_{ac}$ may vary 
from action to action only.  

In our next paper \cite{plaqdual} we derive an exact dual representation of 
non-abelian LGT starting from the plaquette formulation and study in details 
its low-temperature properties. In particular, we shall compute low-temperature 
asymptotics of the dual Boltzmann weight, derive its continuum limit 
and obtain an effective theory for the Wilson loop.

In conclusion, we hope that the present investigation gives a certain background
for an analytical study of gauge models in the low-temperature phase
which is the only phase essential for the construction of the continuum limit.
It can give a solid mathematical basis for conventional PT by proving
(or disproving) its asymptoticity in the large volume limit.
We also think that the present method can give reliable analytical tools
for the investigation of infrared physics relevant for the non-perturbative
phenomena like quark confinement, chiral symmetry breaking, etc.

\begin{appendix}

\section{Properties of the link Green functions}

In this appendix we would like to describe the basic properties of the
link Green functions $G_{ll^{\prime}}$ and $D_l(x^{\prime})$ which
appear to be the  main building blocks of the low-temperature expansion
in the plaquette formulation both in abelian and in non-abelian models.
These functions appear also in the low-temperature expansion of the $2D$ spin
models in the link representation.
They are defined in precisely the same way as in $2D$ and have many properties
analogous to properties of link functions in $2D$, see \cite{2dsunlink} for details.
The functions $G_{ll^{\prime}}$ and $D_l(x^{\prime})$
are defined as\footnote{Please, note that what we call link Green function does not
correspond to link Green function of \cite{pttemp} but rather coincides
up to a Kroneker delta with plaquette Green function of \cite{pttemp}, Eq.(2.24).
We call them link functions only because in $3D$ plaquettes are dual to links.}
\begin{equation}
G_{ll^{\prime}} = 2\delta_{l,l^{\prime}} - G_{x,x^{\prime}} -
G_{x+e_n,x^{\prime}+e_{n^{\prime}}} + G_{x,x^{\prime}+e_{n^{\prime}}} +
G_{x+e_n,x^{\prime}} \ ,
\label{Gll1}
\end{equation}
\noindent
\begin{equation}
D_l(x^{\prime}) = G_{x,x^{\prime}} - G_{x+e_n,x^{\prime}} \ ,
\label{Dxl}
\end{equation}
\noindent
where the link $l=(x,n)$ is defined by a point $x$ and a positive
direction $n$.  $G_{x,x^{\prime}}$ is a ``standard'' Green function on
the periodic (or some other) lattice
\begin{equation}
G_{x,x^{\prime}} = \frac{1}{L^3} \sum_{k_n=0}^{L-1}
\frac{e^{\frac{2\pi i}{L}k_n(x-x^{\prime})_n}}
{f(k)} \ , \ \
f(k)=3-\sum_{n=1}^3\cos \frac{2\pi}{L}k_n \ ,  \ k_n^2\ne 0 \ .
\label{Gxx}
\end{equation}
\noindent
The normalisation is such that $G_{ll}=4/3$.
In the momentum space $G_{ll^{\prime}}$ reads
\begin{equation}
G_{ll^{\prime}} = 2\delta_{l,l^{\prime}} - \frac{1}{L^3}
\sum_{k_n=0}^{L-1}\frac{e^{\frac{2\pi i}{L}k_n(x-x^{\prime})_n}}{f(k)}
(1-e^{\frac{2\pi i}{L}k_n})(1-e^{-\frac{2\pi i}{L}k_{n^{\prime}}}) \ .
\label{Gllmom}
\end{equation}
Using this representation it is easy to prove the following
``orthogonality'' relations for the link functions
\begin{equation}
\sum_{b}G_{lb}G_{bl^{\prime}}=2G_{ll^{\prime}} \ , \
\sum_{b}D_b(x)G_{bl^{\prime}}=0 \ , \
\sum_{b}D_b(x)D_b(x^{\prime})=2G_{x,x^{\prime}} \ ,
\label{orthog}
\end{equation}
\noindent
where the sum over $b$ runs over all links of $3D$ lattice.
Let $\Omega$ be any closed surface. Then
\begin{equation}
\sum_{l,l^{\prime}\in {\Omega}} \bar{G}_{ll^{\prime}} = 0 \ ,
\label{clpath}
\end{equation}
\noindent
where $\bar{G}_{ll^{\prime}}=G_{ll^{\prime}}$ if both link $l$ and $l^{\prime}$
point in either positive or negative direction and
$\bar{G}_{ll^{\prime}}=-G_{ll^{\prime}}$ if one (and only one) of the links
points in negative direction. In particular, Eq.(\ref{clpath}) holds
for each cube of the original lattice.

Let $l_i, i=1,...,6$ be six links attached to a given site $x$ (notations are
as in (\ref{dualln})).
One sees that $G_{ll^{\prime}}$ satisfies the following equation
\begin{equation}
G_{l_1l^{\prime}}+G_{l_2l^{\prime}} + G_{l_3l^{\prime}} - G_{l_4l^{\prime}}-
G_{l_5l^{\prime}} - G_{l_6l^{\prime}} = 0
\label{Glleq}
\end{equation}
\noindent
for any link $l^{\prime}$. $D_l(x^{\prime})$ obeys 
the lattice Laplace equation
\begin{equation}
D_{l_1}(x^{\prime})+D_{l_2}(x^{\prime}) + D_{l_3}(x^{\prime}) -
D_{l_4}(x^{\prime})-D_{l_5}(x^{\prime}) - D_{l_6}(x^{\prime}) =
2\delta_{x,x^{\prime}} \ .
\label{Dleq}
\end{equation}
\noindent
The last three equations ensure the surface independence of the Wilson
loop in the plaquette formulation, they allow to deform some given surface
to any other one.

Let $S^d_{xy}$ be the surface dual to the surface spanned by the Wilson loop.
Consider the Wilson loop $C$ with sizes $R\times T$ and lying in the
$x-y$ plane. Then dual links from the minimal dual surface $S^d_{xy}$ have
the following coordinates $l=(x_1,x_2,x_3;n_3)$, where $x_3$ is arbitrary and
$0\leq x_1 \leq R-1$, $0\leq x_2 \leq T-1$. We then have for the quantity 
$P(C)$ entering the weak-coupling expansion of the Wilson loop (\ref{WLexp})
\begin{equation}
\sum_{l,l^{\prime}\in S^d_{xy}} G_{ll^{\prime}} = 2P(C) \ ,
\label{GLU1}
\end{equation}
\noindent
where
\begin{equation}
P(C) = \frac{1}{L^{3}} \sum_{k_n} \frac{2-\cos p_1-\cos p_2}{f(k)} \
\frac{1-\cos p_1R}{1-\cos p_1} \ \frac{1-\cos p_2T}{1-\cos p_2} \ , \
p_n\equiv\frac{2\pi}{L}k_n \ .
\label{Perim}
\end{equation}
\noindent
In the TL and for $R,T\to\infty$ one finds for $3D$
\begin{equation}
P(C) \approx \frac{1}{\pi}(R\ln T + T\ln R) \ ,
\label{PCTL}
\end{equation}
\noindent
what coincides with \cite{pttemp}.

Finally, we study the asymptotic properties of the link function $G_{ll^{\prime}}$. 
Let $l=(0,x_2,x_3;n_1)$ and $l=(R,x_2,x_3;n_1)$. From the definition 
(\ref{Gll1}) it is easy to obtain
\begin{equation}
G_{ll^{\prime}} = 2D(R) - D(R+1) - D(R-1) \ ,
\label{GllR}
\end{equation}
\noindent
where
\begin{equation}
D(R) = \frac{1}{L^3} \sum_{k_n=0}^{L-1} \
\frac{1-\cos{\frac{2\pi i}{L}k_nR_n}}{f(k)} \ .
\label{DR}
\end{equation}
\noindent
Since when $R\to\infty$
$$
D(R) = \frac{{\rm const}}{R} + {\cal O}(R^{-2})
$$
we come to the following asymptotic behaviour
\begin{equation}
G_{ll^{\prime}} \asymp  \frac{{\rm const}}{R^3} \ .
\label{Gllasymp}
\end{equation}
\noindent

\section{Relation between site and link angles}

Define the site angle $\omega_k(x)$ as follows
\begin{equation}
V_x = \exp \left [ it_k\omega_k(x) \right ] =
\left ( \prod_{i=1}^3e^{it_k\theta_k(l_i)} \right ) e^{it_k\theta_k(C)}
\left ( \prod_{i=4}^6e^{it_k\theta_k(l_i)} \right ) e^{-it_k\theta_k(C)} \ ,
\label{stangldef}
\end{equation}
\noindent
where $V_x\in SU(N)$, $t_k$ are the generators of the $SU(N)$ group and
\begin{equation}
e^{it_k\theta_k(C)} = \prod_{b\in C}e^{it_k\theta_k(b)} \ .
\label{conangldef}
\end{equation}
\noindent
$\omega_k(x)$ has an expansion in powers of the link angles
$\theta_k(l)$ which can be obtained by making repeated use of
the Campbell-Baker-Hausdorf formula. The expansion is of the form
\begin{equation}
\omega_k(x) =  \omega^{(0)}_k(x) +
\omega^{(1)}_k(x,C) + \omega^{(2)}_k(x,C) + \cdots \ ,
\label{stanglexp}
\end{equation}
\noindent
where we have indicated explicitly the dependence of the coefficients
on the connector $C$. Let us denote
\begin{equation}
\theta^{(0)}_k(C) = \sum_{b\in C} \theta_k(b) \ .
\label{conanglexp}
\end{equation}
\noindent
Then the first three terms in the expansion (\ref{stanglexp}) have the
following form
\begin{equation}
\omega^{(0)}_k(x) = \sum_{i=1}^6 \theta_k(l_i) \ ,
\label{wkx0app}
\end{equation}
\noindent
\begin{equation}
\omega^{(1)}_k(x,C) = \omega^{(1)}_k(x) + \omega^{(1)}_k(C) \ ,
\label{wkx1}
\end{equation}
\noindent
\begin{equation}
\omega^{(2)}_k(x,C) = \omega^{(2)}_k(x) + \omega^{(2)}_k(C) \ ,
\label{wkx2}
\end{equation}
\noindent
where we have separated the connector contributions explicitly, i.e.
$\omega^{(i)}_k(x)$ depend only on six links $l_i$ which belong to
a given site $x$, while $\omega^{(i)}_k(C)$ depend also on links
$b\in C$. All these coefficients can be written down as
\begin{equation}
\omega^{(1)}_k(x) = -\epsilon_{kmn}\sum_{i<j}^6
\theta_m(l_i)\theta_n(l_j) \ ,
\label{wkx1site}
\end{equation}
\noindent
\begin{equation}
\omega^{(1)}_k(C) = -2\epsilon_{kmn}\theta_m^{(0)}(C)\sum_{i=4}^6
\theta_n(l_i) \ ,
\label{wkx1con}
\end{equation}
\noindent
\begin{eqnarray}
\omega^{(2)}_k(x) = \frac{1}{3}\sum_{ij=1}^6 \left [ \theta_k(l_i)
\theta_n(l_i)\theta_n(l_j) - \theta_n(l_i)\theta_n(l_i)\theta_k(l_j) \right ]
\nonumber \\
+ \frac{2}{3}\sum_{i<j<p}^6 \left [ 2\theta_n(l_i)
\theta_k(l_j)\theta_n(l_p) - \theta_n(l_i)\theta_n(l_j)\theta_k(l_p) -
\theta_k(l_i)\theta_n(l_j)\theta_n(l_p) \right ] \ ,
\label{wkx2site}
\end{eqnarray}
\noindent
\begin{eqnarray}
\omega^{(2)}_k(C) = 2\sum_{i=1}^3\sum_{j=4}^6 \left [ \theta_n(l_i)\theta_n(l_j)
\theta_k^{(0)}(C) - \theta_n(l_i)\theta_k(l_j)\theta_n^{(0)}(C) \right ]
\nonumber \\
+ 2\sum_{3<i<j}^6 \left [ \theta_k(l_i)
\theta_n(l_j)\theta_n^{(0)}(C) - \theta_n(l_i)\theta_k(l_j)\theta_n^{(0)}(C) \right ]
\nonumber  \\
+ 2\sum_{i=4}^6 \left [ \sum_{b\in C}\theta_k(b)
\theta_n(b)\theta_n(l_i) +2 \sum_{b<b^{\prime}\in C}\theta_n(b)
\theta_k(b^{\prime})\theta_n(l_i) - \theta_n^{(0)}(C)\theta_n^{(0)}(C)
\theta_k(l_i) \right ]  .
\label{wkx2con}
\end{eqnarray}
\noindent

\section{Exact expression for the first coefficient of the free energy 
expansion of $SU(N)$ model}

Here we present the computation of the first order coefficient of
the free energy expansion of the $SU(N)$ model
\begin{equation}
F=\frac{1}{3NL^3}\ln Z = \beta - \frac{N^2-1}{3N}\ln\beta - 
\sum_{f=1}^{N_f}\frac{{\rm Tr}\ln \hat{G}_f}{3NL^3} + 
\frac{1}{3N\beta L^3}C_1 + {\cal O}(\beta^{-2}) \ .
\label{fren}
\end{equation}
\noindent
There are five contributions to $C_1$
\begin{equation}
C_1 = C_{ac} + C_{meas} + C_{J1} +  C_{J2} + C_{ferm} \ .
\label{frensum}
\end{equation}
\noindent
The contribution from the action is given by
\begin{equation}
\frac{1}{3NL^3}C_{ac}=\frac{(N^2-1)(2N^2-3)}{96N^2} \
\frac{1}{3L^3} \ \sum_lG^2_{ll} \ = \ \frac{(N^2-1)(2N^2-3)}{54N^2} 
\label{c1ac}
\end{equation}
\noindent
and the contribution from the measure by
\begin{equation}
\frac{1}{3NL^3}C_{meas}= - \frac{N^2-1}{24} \
\frac{1}{3L^3} \ \sum_lG_{ll} \ =  \ - \frac{N^2-1}{18} \ .
\label{c1meas}
\end{equation}
\noindent
As is seen from the first line in the Eq.(\ref{PFwkexp}) there are two
contributions from the expansion of the Jacobian. They are given by
the following expectation values
\begin{equation}
\frac{1}{3NL^3}C_{J1} = \frac{1}{3NL^3}(-\frac{i}{2})\sum_{x,k} \
\left \langle \alpha_k(x)\omega_k^{(2)}(x) \right \rangle \ ,
\label{CJ1def}
\end{equation}
\noindent
\begin{equation}
\frac{1}{3NL^3}C_{J2} = \frac{1}{3NL^3}(-\frac{1}{4}) \
\left \langle \left ( \sum_{x,k}\alpha_k(x)\omega_k^{(1)}(x)
\right )^2 \right \rangle \ .
\label{CJ2def}
\end{equation}
\noindent
To compute these expectation values we have to use the representation of
site angles $\omega_k^{(i)}(x)$ in terms of link angles $\omega_k(l)$ given
in Appendix B, where $\theta_k(l)$ coincide with $\omega_k(l)$ up to
a sign. Therefore, we can define
\begin{equation}
\omega_k(l) = \eta_l \theta_k(l) \ , \ \eta_l = \pm 1 \ .
\label{linksign}
\end{equation}
\noindent
This sign $\eta_l$ can be easily established for any link from the explicit
expressions for the Bianchi identities given in Eqs.(\ref{V1})-(\ref{CB4A}).
Let $l_i$, $i=1,...,6$ denote six links attached to a site $x(c)$ from
the $c$-th BCC sub-lattice. Let $C$ be a set of links $b$ which form
a connector for a given site $x(c)$. Introduce now the following quantities
\begin{eqnarray}
M_1(x(c);l_1l_2l_3) = \frac{1}{3}\sum_{ij=1}^6\eta_{l_j}
\left [ \delta_{l_il_1} \delta_{l_il_2} \delta_{l_jl_3} -
\delta_{l_il_2} \delta_{l_il_3} \delta_{l_jl_1}  \right ] \nonumber  \\
+ \frac{2}{3}\sum_{i<j<p}^6\eta_{l_i}\eta_{l_j}\eta_{l_p}
\left [ 2\delta_{l_il_2} \delta_{l_jl_1} \delta_{l_pl_3} -
\delta_{l_il_2} \delta_{l_jl_3} \delta_{l_pl_1} -
\delta_{l_il_1} \delta_{l_jl_2} \delta_{l_pl_3} \right ] \ ,
\label{M1}
\end{eqnarray}
\noindent
\begin{eqnarray}
M_2(x(c);l_1l_2l_3) = 2\sum_{i=1}^3\sum_{j=4}^6\eta_{l_i}\eta_{l_j}
\sum_{b\in C}\eta_b
\left [ \delta_{l_il_2} \delta_{l_jl_3} \delta_{bl_1} -
\delta_{l_il_2} \delta_{l_jl_1} \delta_{bl_3}  \right ] \nonumber  \\
+ 2\sum_{3<i<j}^6\eta_{l_i}\eta_{l_j}\sum_{b\in C}\eta_b
\left [ \delta_{l_il_1} \delta_{l_jl_2} \delta_{bl_3} -
\delta_{l_il_2} \delta_{l_jl_1} \delta_{bl_3} \right ] \nonumber  \\
+2\sum_{i=4}^6 \left [ \sum_{b\in C}\eta_{l_i}
\delta_{bl_1} \delta_{bl_2} \delta_{l_il_3} + 2 \sum_{b<b^{\prime}\in C}
\eta_{l_i}\eta_b\eta_{b^{\prime}} \delta_{bl_2} \delta_{b^{\prime}l_1}
\delta_{l_il_3} - \sum_{b,b^{\prime}\in C} \eta_{l_i}\eta_b\eta_{b^{\prime}}
\delta_{bl_2} \delta_{b^{\prime}l_3} \delta_{l_il_1} \right ] ,
\label{M2}
\end{eqnarray}
\noindent
\begin{equation}
B(x(c),x^{\prime}(c^{\prime});l_1l_2) =
B(x(c);l_1) B(x^{\prime}(c^{\prime});l_2) \ ,
\label{Bdef}
\end{equation}
\noindent
\begin{equation}
B(x(c);l_1) =
\left [ \sum_{i<j}^6\eta_{l_i}\eta_{l_j}\delta_{l_il_1} +
2\sum_{b\in C}\eta_b\sum_{j=4}^6\eta_{l_j}\delta_{bl_1} \right ] \ .
\label{Bx}
\end{equation}
\noindent

With the help of these quantities, and after taking all the derivatives and 
calculating the sums over the group indices we present the result in the form
\begin{equation}
\frac{1}{3NL^3}C_{J1} = \frac{N^2-1}{2N} \ \left [ A_1 + A_2 \right ] \ ,
\label{CJ1res}
\end{equation}
\noindent
\begin{equation}
\frac{1}{3NL^3}C_{J2} = \frac{(N^2-1)(N^2-2)}{32N} \
\left [ Q_1 + Q_2 \right ] \ ,
\label{CJ2ref}
\end{equation}
\noindent
where
\begin{eqnarray}
A_i = \frac{1}{3L^3}\sum_{c=1}^4 \ \sum_{x(c)} \
\sum_{l_1l_2l_3}M_i(x(c);l_1l_2l_3) \nonumber  \\
\left [ (N^2-1)D_{l_1}(x(c))G_{l_2l_3} + D_{l_2}(x(c))G_{l_1l_3} +
D_{l_3}(x(c))G_{l_2l_1} \right ] \ ,
\label{A1def}
\end{eqnarray}
\noindent
\begin{equation}
\label{Q1def}
Q_1 = \frac{1}{3L^3}\sum_{c,c^{\prime}=1}^4\sum_{x(c),x^{\prime}(c^{\prime})}
\sum_{l_1l_2} B(x(c),x^{\prime}(c^{\prime});l_1l_2)
[ G_{l_1,l^{\prime}_{j^{\prime}}}D_{l_2}(x(c)) D_{l_j}(x^{\prime}(c^{\prime})) +
\end{equation}
$$
G_{l_j,l_2}D_{l^{\prime}_{j^{\prime}}}(x(c))
D_{l_1}(x^{\prime}(c^{\prime})) -
G_{l_1,l_2}D_{l^{\prime}_{j^{\prime}}}(x(c))
D_{l_{j}}(x^{\prime}(c^{\prime})) -
G_{l_j,l^{\prime}_{j^{\prime}}}D_{l_2}(x(c))
D_{l_1}(x^{\prime}(c^{\prime})) ] \ ,
$$
\noindent
\begin{equation}
Q_2 = \frac{1}{3L^3}\sum_{c,c^{\prime}=1}^4\sum_{x(c),x^{\prime}(c^{\prime})}
\sum_{l_1l_2} B(x(c),x^{\prime}(c^{\prime});l_1l_2)
G_{x(c),x^{\prime}(c^{\prime})}
\left [ G_{l_1,l^{\prime}_{j^{\prime}}}G_{l_j,l_2} -
G_{l_1,l_2}G_{l_j,l^{\prime}_{j^{\prime}}} \right ] \ .
\label{Q2def}
\end{equation}
\noindent

Finally, we compute the contribution of fermions. As follows from 
(\ref{fermmatrinteract})-(\ref{fermptexp}) two $1/\beta$-contributions 
are given by 
\begin{equation}
C_{ferm} = -\frac{N}{8}\sum_f\sum_pH_f(p)\langle \omega_k^2(p) \rangle 
+ \frac{N}{16}\sum_f\sum_{pp^{\prime}}H_f(pp^{\prime})
\langle \omega_k(p)\omega_k(p^{\prime}) \rangle  \ . 
\label{Cferm}
\end{equation}
\noindent
We thus obtain 
\begin{equation}
\frac{1}{3NL^3}C_{ferm} = \frac{N^2-1}{48L^3}\sum_{f=1}^{N_f} V_f \ ,  
\label{Cfermfin}
\end{equation}
\begin{equation}
V_f = \left [ - \sum_pH_f(p)\sum_{(l,l^{\prime})\in p} + 
\frac{1}{2}\sum_{pp^{\prime}}H_f(pp^{\prime})\sum_{l\in p} 
\sum_{l^{\prime}\in p^{\prime}} \right ] G_{ll^{\prime}} \ , 
\label{Vferm}
\end{equation}
where we have denoted 
\begin{equation}
H_f(p) = {\rm Tr}_s\left [ \hat{G}_f(c+e_n,c)\xi_n(c) +
\hat{G}_f(c,c+e_n)\overline{\xi}_n(c)  \right ] \ ,
\label{Hfp}
\end{equation}
\begin{eqnarray}
H_f(pp^{\prime}) &=& {\rm Tr}_s \ [ \ \hat{G}_f(c^{\prime}+e_{n^{\prime}},c)\xi_n(c)
\hat{G}_f(c+e_n,c^{\prime})\xi_{n^{\prime}}(c^{\prime}) 
\nonumber   \\
&+& \hat{G}_f(c^{\prime},c+e_n)\overline{\xi}_n(c)
\hat{G}_f(c,c^{\prime}+e_{n^{\prime}})\overline{\xi}_{n^{\prime}}(c^{\prime})  
\nonumber   \\
&-& \hat{G}_f(c^{\prime},c)\xi_n(c) \hat{G}_f(c+e_n,c^{\prime}+e_{n^{\prime}})
\overline{\xi}_{n^{\prime}}(c^{\prime}) 
\nonumber   \\
&-& \hat{G}_f(c^{\prime}+e_{n^{\prime}},c+e_n)\overline{\xi}_n(c) 
\hat{G}_f(c,c^{\prime})\xi_{n^{\prime}}(c^{\prime}) ] 
\label{Hfpp2}
\end{eqnarray}
and ${\rm Tr}_s$ means the trace over spinor indices. 

\end{appendix}


\begin{thebibliography}{99}


\bibitem{wilson} K.~G.~Wilson, Phys.Rev. D10 (1974) 2445.
%
\bibitem{dualu1} T.~Banks, J.~Kogut, R.~Myerson, Nucl.Phys. B121 (1977) 493.
%
\bibitem{dualzn} M.~Einhorn,  R.~Savits, Phys.Rev. D17 (1978) 2583.
%
\bibitem{dualsun} A.~Ukawa, P.~Windey, A.~Guth, Phys.Rev. D21 (1980) 1013.
%
\bibitem{india} R.~Anishetty, S.~Cheluvaraja,
H.S.~Sharatchandra and M.~Mathur, Phys.Lett. B314 (1993) 387.
%
\bibitem{dual4d} I.~Halliday, P.~Suranyi, Phys.Lett. B350 (1995) 189.
%
\bibitem{camb} R.~Oeckl, H.~Pfeiffer, Nucl.Phys. B598 (2001) 400-426.
%
\bibitem{dkpt} D.~Diakonov, V.~Petrov, Journal Exp. Theor. Phys. 
91 (2000) 873-893.
%
\bibitem{crimea} O.~Borisenko, M.~Faber, Dual representation for lattice 
gauge models, Proc. of the International
School-Conference ``New trends in high-energy physics'',
Ed.by P.~Bogolyubov, L.~Jenkovszky, Kiev (2000) 221.
%
\bibitem{conf4} O.~Borisenko, M.~Faber, Confinement picture in dual
formulation of lattice gauge models, Proc. of the Vienna International
Symposium ``Confinement-IV'', 2001, World Scientific Publishing,
Singapore-New-Jersey-London-Hong-Kong, 269.
%
\bibitem{halpern} M.B.~Halpern, Phys.Rev. D19 (1979) 517;
Phys.Lett. B81 (1979) 245.
%
\bibitem{plrepr} G.~Batrouni, Nucl.Phys. B208 (1982) 467.
%
\bibitem{u1dec} A.~Guth, Phys.Rev. D21 (1980) 2291.
%
\bibitem{rigu1lgt} J.~Fr\"ohlich, T.~Spencer, Commun.Math.Phys. 83 (1982) 411.
%
\bibitem{mack} M.~G\"{o}pfert, G.~Mack, Commun.Math.Phys. 81 (1981) 97;
82 (1982) 545.
%
\bibitem{zachdiss} M.~Zach, M.~Faber, P.~Skala, Nucl.Phys. B529 (1998) 505;
Phys.Rev. D57 (1998) 123.
%
\bibitem{rusakov} B.~Rusakov, Phys.Lett. B398 (1997) 331; Nucl.Phys. B507 (1997) 691.
%
\bibitem{ginvplaq} G.~Batrouni, M.B.~Halpern, Phys.Rev. D30 (1984) 1782.
%
\bibitem{meanplaq} G.~Batrouni, Nucl.Phys. B208 (1982) 12.
%
\bibitem{mackpetk} G.~Mack, V.~Petkova, Ann.Phys. 125 (1980) 117.
%
\bibitem{seiler} A.~Patrascioiu, E.~Seiler, Phys.Rev.Lett.
74 (1995) 1924.
%
\bibitem{plaqform} O.~Borisenko, S.~Voloshin, M.~Faber, Analytical study of 
low temperature phase of $3D$ LGT in the plaquette formulation,  
Proc. of NATO Workshop "Confinement, Topology and Other Non-perturbative Aspects 
of QCD", Ed. by J.~Greensite, and S.~Olejnik, Kluwer Academic Publishers, 2002, 33.
%
\bibitem{plaqferm} O.~Borisenko, S.~Voloshin, Field-strength formulation 
of lattice QCD with dynamical fermions and related topological structure, 
Proceedings of XVI International Symposium ISHEPP, Dubna, Russia, 2002.
%
\bibitem{2dsunlink} O.~Borisenko, V.~Kushnir, A.~Velytsky, Phys.Rev. D62 (2000) 025013;
e-print archive hep-lat/9809133, hep-lat/9905025.
%
\bibitem{linkrepr} G.~Batrouni, M.B.~Halpern, Phys.Rev. D30 (1984) 1775.
%
\bibitem{pttemp} V.F.~M\"uller, W.~R\"uhl, Ann.Phys. 133 (1981) 240.
%
\bibitem{plaqdual} O.~Borisenko, S.~Voloshin, M.~Faber,  Plaquette representation 
for $3D$ lattice gauge models: II. Dual form and its properties, in preparation.


\end{thebibliography}
\end{document}